\documentclass[preprint,showpacs,preprintnumbers,amsmath,amssymb]{revtex4}
\usepackage{graphicx}
\usepackage{dcolumn}
\usepackage{bm}

\begin{document}

\title{Implementation of Time-Delay Interferometry for LISA}

\author{Massimo Tinto} 
\email{Massimo.Tinto@jpl.nasa.gov}
\altaffiliation [Also at: ]{Space Radiation Laboratory, California
  Institute of Technology, Pasadena, CA 91125} 
\affiliation{Jet Propulsion Laboratory, California 
Institute of Technology, Pasadena, CA 91109}

\author{Daniel A. Shaddock}
\email{Daniel.A.Shaddock@jpl.nasa.gov}
\affiliation{Jet Propulsion Laboratory, California 
Institute of Technology, Pasadena, CA 91109}

\author{Julien Sylvestre} 
\email{jsylvest@ligo.caltech.edu}
\altaffiliation [Also at: ]{LIGO Laboratory, California
Institute of Technology, Pasadena, CA 91125}
\affiliation{Jet Propulsion Laboratory, California 
Institute of Technology, Pasadena, CA 91109}

\author{J.W. Armstrong}
\email{John.W.Armstrong@jpl.nasa.gov}
\affiliation{Jet Propulsion Laboratory, California Institute of Technology,
 Pasadena, CA 91109}

\date{\today}
\vskip24pt

\begin{abstract}
  We discuss the baseline optical configuration for the Laser
  Interferometer Space Antenna (LISA) mission, in which the lasers are
  not free-running, but rather one of them is used as the main
  frequency reference generator (the {\it master}) and the remaining
  five as {\it slaves}, these being phase-locked to the master (the
  {\it master-slave configuration}).  Under the condition that the
  frequency fluctuations due to the optical transponders can be made
  negligible with respect to the secondary LISA noise sources (mainly
  proof-mass and shot noises), we show that the entire space of
  interferometric combinations LISA can generate when operated with
  six independent lasers (the {\it one-way method}) can also be
  constructed with the {\it master-slave} system design.  The
  corresponding hardware trade-off analysis for these two optical
  designs is presented, which indicates that the two sets of systems
  needed for implementing the {\it one-way method}, and the {\it
    master-slave configuration}, are essentially identical.  Either
  operational mode could therefore be implemented without major
  implications on the hardware configuration.
  
  We then derive the required accuracies of armlength knowledge, time
  synchronization of the onboard clocks, sampling times and
  time-shifts needed for effectively implementing Time-Delay
  Interferometry for LISA.  We find that an armlength accuracy of
  about $16$ meters, a synchronization accuracy of about $50$ ns, and
  the time jitter due to a presently existing space qualified clock
  will allow the suppression of the frequency fluctuations of the
  lasers below to the level identified by the secondary noise sources.
  A new procedure for sampling the data in such a way to avoid the
  problem of having time shifts that are not integer multiples of the
  sampling time is also introduced, addressing one of the concerns
  about the implementation of Time-Delay Interferometry.
\end{abstract}


\maketitle

\section{Introduction}

LISA, the Laser Interferometer Space Antenna, is a three-spacecraft
deep space mission, jointly proposed to the National Aeronautics and
Space Administration (NASA) and the European Space Agency (ESA). The
LISA scientific objective is to detect and study low-frequency cosmic
gravitational radiation by observing phase differences of laser beams
interchanged between drag-free spacecraft \cite{1}.

Modeling each spacecraft with two optical benches, carrying
independent lasers, frequency generators (called Ultra Stable
Oscillators), beam splitters and photo receivers, the measured
eighteen time series of frequency shifts (six obtained from the six
one-way laser beams between spacecraft pairs, six from the beams
between the two optical benches on each of the three spacecraft, and
six more from modulation of the laser beams with USO data) were
previously analyzed by Tinto {\it et al.} \cite{2}.  There it was
shown that there exist several combinations of these eighteen
observables which exactly cancel the otherwise overwhelming phase noise
of the lasers, the phase fluctuations due to the non-inertial motions
of the six optical benches, and the phase fluctuations introduced by
the three Ultra Stable Oscillators into the heterodyned measurements,
while leaving effects due to passing gravitational waves.

The analysis presented in \cite{2} relied on the assumptions that (i)
the frequency offsets of any pair of independent lasers (assumed there
to be $ \approx \ 300 \ {\rm MHz}$) could be observed within the
detection bandwidths of the photo receivers where the one-way Doppler
measurements are performed, and (ii) the telemetry data rate needed by
two of the three spacecraft to transmit their measured one-way Doppler
data to the third spacecraft (where the interferometric combinations
are synthesized) is adequate.  Although the technology LISA will be
able to use should make possible the implementation of Time-Delay
Interferometry (TDI) as discussed in \cite{2}, the possibility of
optimizing the design of the optical layout, while at the same time
minimizing the number of Doppler data needed for constructing the
entire space of interferometric observables, was not analyzed there.
Here we extend those results to a different optical configuration, in
which one of the six lasers is the provider of the frequency reference
(albeit time-delayed) for the other five via phase-locking.  This {\it
  master-slave} optical design could provide potential advantages,
such as smaller frequency offsets between beams from pairs of
different lasers, hardware redundancy, reliability, and can result in
a smaller number of measured data. An outline of the paper is given
below.

In Section II we summarize TDI, the data processing technique needed
for removing the frequency fluctuations of the six lasers used by
LISA, and other noises.  In order to show that the entire space of
interferometric observables LISA can generate can also be
reconstructed by using a master-slave optical configuration, we
consider the simple case of spacecraft that are stationary with
respect to each other. After showing that the entire space of
interferometric observables can be obtained by properly combining four
generators, ($\alpha, \beta, \gamma, \zeta$), we then derive the
expressions for the four generators corresponding to the master-slave
optical configuration.  By imposing some of the one-way measurements
entering into ($\alpha, \beta, \gamma, \zeta$) to be zero (the so
called {\it locking conditions}), we show that the expressions for
these generators can be written in terms of the one-way and two-way
Doppler measurements corresponding to the locking configuration we
analyzed. Section III and Appendix A provide a theoretical derivation
and estimation of the magnitude of the phase noise expected to be
generated by an optical transponder. In Section IV we analyze and
compare the hardware requirements needed for implementing both optical
designs, while in Section V we turn to the estimation of the armlength
and time synchronization accuracies, as well as time-shift and
sampling time precisions needed for successfully implementing TDI with
LISA.  Our comments and conclusions are finally presented in Section
VI.

\section{Time-Delay Interferometry}

The description of TDI for LISA is greatly simplified if we adopt the
notation shown in Figure 1, where the overall geometry of the LISA
detector is defined. The spacecraft are labeled $1$, $2$, $3$ and
distances between pairs of spacecraft are $L_1$, $L_2$, $L_3$, with
$L_i$ being opposite spacecraft $i$.  Unit vectors between spacecraft
are $\hat n_i$, oriented as indicated in figure 1.  We similarly index
the phase difference data to be analyzed: $s_{31}$ is the phase
difference time series measured at reception at spacecraft 1 with
transmission from spacecraft 2 (along $L_3$).
\begin{figure}
\centering
\includegraphics[width=2.5in, angle=-90]{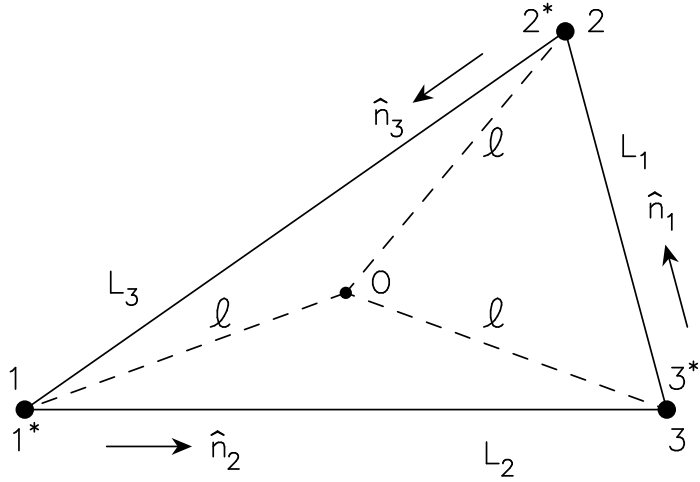}
\caption{Schematic LISA configuration.  Each spacecraft is equidistant
from the point O, in the plane of the spacecraft.  Unit vectors
$\hat n_i$ point between spacecraft pairs with the indicated
orientation.  At each vertex spacecraft there are two optical
benches (denoted 1, $1^*$, etc.), as indicated.}
\end{figure}
Similarly, $s_{21}$ is the phase difference series derived from
reception at spacecraft $1$ with transmission from spacecraft $3$. The
other four one-way phase difference time series from signals exchanged
between the spacecraft are obtained by cyclic permutation of the
indices: $1$ $\to$ $2$ $\to$ $3$ $\to$ $1$.  We also adopt a useful
notation for delayed data streams: $s_{31,2} = s_{31}(t - L_2)$,
$s_{31,23} = s_{31}(t - L_2 - L_3) = s_{31,32}$, etc.  (we take the
speed of light $c = 1$ for the analysis).  Six more phase difference
series result from laser beams exchanged between adjacent optical
benches within each spacecraft; these are similarly indexed as
$\tau_{ij}$ ($i,j = 1, 2, 3 \ ; \ i \ne j$).

The proof-mass-plus-optical-bench assemblies for LISA spacecraft
number $1$ are shown schematically in figure 2.  We take the left-hand
optical bench to be bench number $1$, while the right-hand bench is
$1^*$.  The photo receivers that generate the data $s_{21}$, $s_{31}$,
$\tau_{21}$, and $\tau_{31}$ at spacecraft $1$ are shown.  The phase
fluctuations of the laser on optical bench $1$ is $p_1(t)$; on optical
bench $1^*$ it is $p^*_1(t)$ and these are independent (the lasers are
for the moment not "locked" to each other, and both are referenced to
their own independent frequency stabilizing device).  We extend the cyclic
terminology so that at vertex $i$ ($i = 1, 2, 3$) the random displacement
vectors of the two proof masses are respectively denoted $\vec
\delta_i(t)$ and $\vec \delta^*_i(t)$, and the random displacements
(perhaps several orders of magnitude greater) of their optical benches
are correspondingly denoted $\vec \Delta_i(t)$ and $\vec
\Delta^*_i(t)$.  As pointed out in \cite{3}, the analysis does
\underline {not} assume that pairs of optical benches are rigidly
connected, i.e. $\vec \Delta_i \neq \vec \Delta^*_i$, in general.  The
present LISA design shows optical fibers transmitting signals both
ways between adjacent benches.  We ignore time-delay effects for these
signals and will simply denote by $\mu_i(t)$ the phase fluctuations
upon transmission through the fibers of the laser beams with
frequencies $\nu_{i}$, and $\nu^*_{i}$.  The $\mu_i (t)$ phase shifts
within a given spacecraft might not be the same for large frequency
differences $\nu_{i} - \nu^*_{i}$. For the envisioned frequency
differences (a few hundred megahertz), however, the remaining
fluctuations due to the optical fiber can be neglected \cite{4}.  It
is also assumed that the phase noise added by the fibers is
independent of the direction of light propagation through them.

Figure 2 endeavors to make the detailed light paths for these
observations clear.  An outgoing light beam transmitted to a distant
spacecraft is routed from the laser on the local optical bench using
mirrors and beam splitters; this beam does not interact with the local
proof mass.  Conversely, an {\it {incoming}} light beam from a distant
spacecraft is bounced off the local proof mass before being reflected
onto the photo receiver where it is mixed with light from the laser on
that same optical bench. The inter-spacecraft phase data are denoted
$s_{31}$ and $s_{21}$ in figure 2.
\begin{figure}
\centering
\includegraphics[width=6.0 in, angle=0]{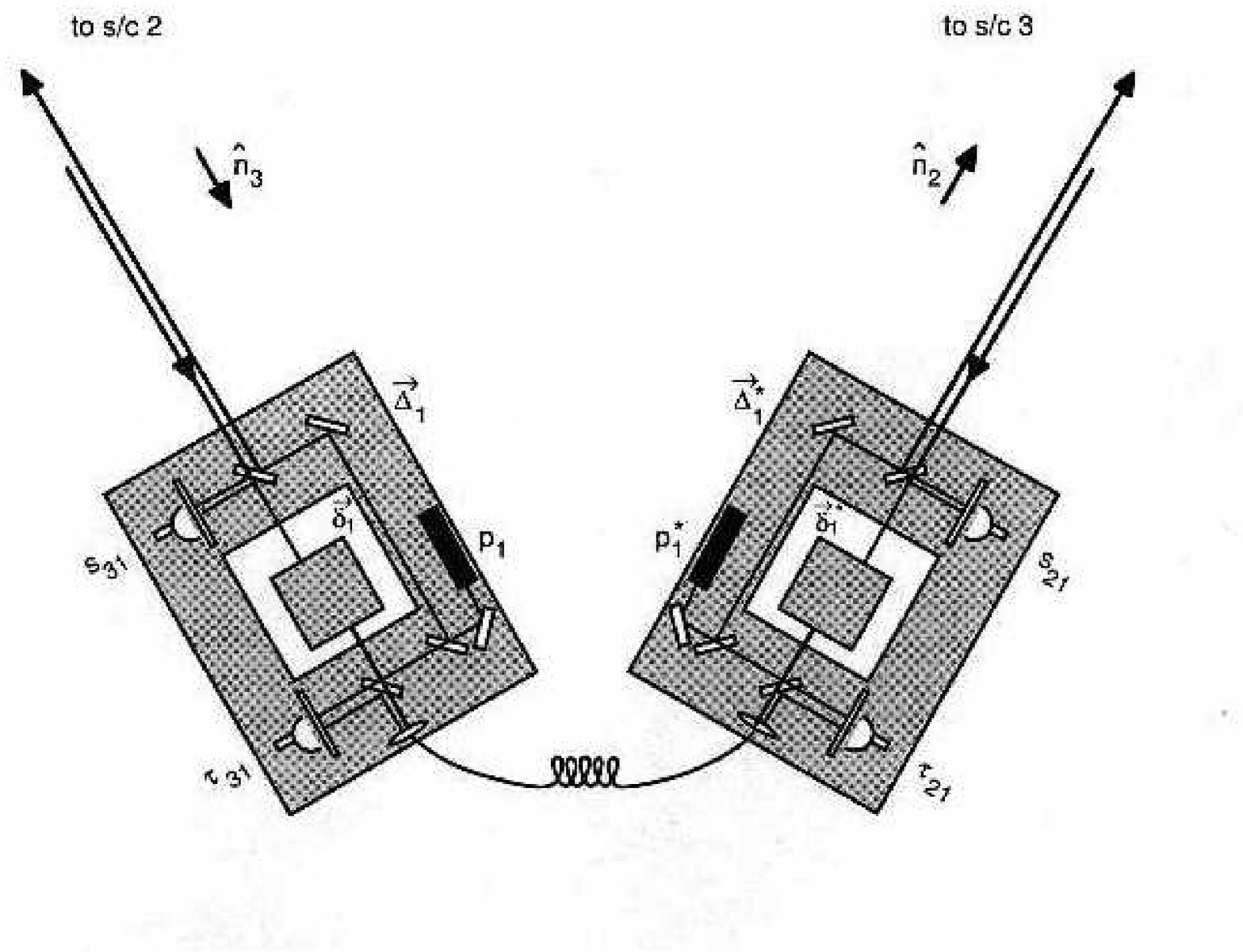}
\caption{Schematic diagram of proof-masses-plus-optical-benches
  for a LISA spacecraft.  The left-hand bench, $1$, reads out the
  phase signals $s_{31}$ and $\tau_{31}$.  The right hand bench,
  $1^*$, analogously reads out $s_{21}$ and $\tau_{21}$.  The random
  displacements of the two proof masses and two optical benches are
  indicated (lower case $\vec \delta_i$ for the proof masses, upper
  case $\vec \Delta_i$ for the optical benches.)}
\end{figure}
Beams between adjacent optical benches within a single spacecraft are
bounced off proof masses in the opposite way.  Light to be {\it
  {transmitted}} from the laser on an optical bench is {\it {first}}
bounced off the proof mass it encloses and then directed to the other
optical bench.  Upon reception it does {\it not} interact with the
proof mass there, but is directly mixed with local laser light, and
again down converted. These data are denoted $\tau_{31}$ and
$\tau_{21}$ in figure 2.

The terms in the following equations for the $s_{ij}$ and $\tau_{ij}$
phase measurements can now be developed from figures 1 and 2, and they
are for the particular LISA configuration in which all the lasers have
the same nominal frequency $\nu_0$, and the spacecraft are stationary
with respect to each other. The analysis covering the configuration
with lasers of different frequencies and spacecraft moving relative to
each other was done in \cite{2}, and we refer the reader to that
paper.

Consider the $s_{31} (t)$ process (equation (\ref{eq:3})) below.  The
photo receiver on the left bench of spacecraft $1$, which (in the
spacecraft frame) experiences a time-varying displacement $\vec
\Delta_1$, measures the phase difference $s_{31}$ by first mixing the
beam from the distant optical bench $2^*$ in direction $\hat n_3$, and
laser phase noise $p^*_2$ and optical bench motion $\vec \Delta^*_2$
that have been delayed by propagation along $L_3$, after one bounce
off the proof mass ($\vec \delta_1$), with the local laser light (with
phase noise $p_1$). Since for this simplified configuration no
frequency offsets are present, there is of course no need for any
heterodyne conversion. 

In equation (\ref{eq:4}) the $\tau_{31}$ measurement results from
light originating at the right-bench laser ($p^*_1$, $\vec
\Delta^*_1$), bounced once off the right proof mass ($\vec
\delta^*_1$), and directed through the fiber (incurring phase shift
$\mu_1(t)$), to the left bench, where it is mixed with laser light
($p_1$).  Similarly the right bench records the phase differences
$s_{21}$ and $\tau_{21}$.  The laser noises, the gravitational wave
signals, the optical path noises, and proof-mass and bench noises,
enter into the four data streams recorded at vertex
$1$ according to the following expressions \cite{2}
\begin{eqnarray}
s_{21} & = &  s^{\rm gw}_{21} + s^{\rm opt. \ path}_{21} +
p_{3,2} - p^{*}_1 + \nu_{0} \ \left[ 2 {\hat n_2} \cdot {\vec \delta^{*}_1}  - 
{\hat n_2} \cdot {\vec \Delta^{*}_1} - {\hat n_2} \cdot {\vec
  \Delta_{3,2}} \right] \ ,
\label{eq:1}
\\
\tau_{21} & = &  p_{1} - p^{*}_1 + 2 \ \nu_{0} \ {\hat n_3} \cdot 
({\vec \delta_1} - {\vec \Delta_1}) + \mu_1 \ ,
\label{eq:2}
\\
s_{31} & = & s^{\rm gw}_{31} + s^{\rm opt. \ path}_{31} + 
p^*_{2,3} - p_1 + \nu_{0} \ \left[ - \ 2 {\hat n_3} \cdot {\vec \delta_1} +
{\hat n_3} \cdot {\vec \Delta_1} + {\hat n_3} \cdot {\vec
  \Delta^*_{2,3}} \right] \ ,
\label{eq:3}
\\
\tau_{31} & = & p^{*}_{1} - p_1 
- \ 2 \ \nu_{0} \ {\hat n_2} \cdot ({\vec \delta^*_1} - {\vec
  \Delta^*_1}) + \mu_1 \ .
\label{eq:4}
\end{eqnarray}
\noindent
Eight other relations, for the readouts at vertices 2 and 3, are given
by cyclic permutation of the indices in equations
(\ref{eq:1})-(\ref{eq:4}).

The gravitational wave phase signal components, $s^{\rm gw}_{ij} \ , \
i, j = 1, 2, 3$, in equations (\ref{eq:1}) and (\ref{eq:3}) are
given by integrating with respect to time the equations (1), (2) of
reference \cite{5} that relate metric perturbations to frequency
shifts. The optical path phase noise contributions, $s^{\rm opt. \ 
  path}_{ij}$, which includes shot noise from the low signal-to-noise
ratio (SNR) in the links between the distant spacecraft, can be
derived from the corresponding term given in \cite{3}. The $\tau_{ij}$
measurements will be made with high SNR so that for them the shot
noise is negligible.

The laser-noise-free combinations of phase data can readily be
obtained from those given in \cite{3} for frequency data.  We use the
same notations: $X$, $Y$, $Z$, $\alpha$, $\beta$, $\gamma$, $\zeta$,
etc., but the reader should keep in mind that here these are phase
measurements.

The phase fluctuations, $s_{ij} \ , \ \tau_{ij} \ , \ i,j=1, 2, 3$,
are the fundamental measurements needed to synthesize all the
interferometric observables unaffected by laser and optical bench
noises.  If we assume for the moment these phase measurements to be
continuous functions of time, the three armlengths to be perfectly
known and constant, and the three clocks onboard the spacecraft to be
perfectly synchronized, then it is possible to cancel out exactly the
phase fluctuations due to the six lasers and six optical benches by
properly time-shifting and linearly combining the twelve measurements
$s_{ij} \ , \ \tau_{ij} \ , \ i,j=1, 2, 3$.  The simplest such
combination, the totally symmetrized Sagnac response $\zeta$, uses all
the data of Figure 2 symmetrically
\begin{eqnarray}
\zeta & = & s_{32,2} - s_{23,3} + s_{13,3} - s_{31,1} + s_{21,1} - s_{12,2}
\nonumber \\
& & + {1 \over 2}
[ ({\tau_{23}} - {\tau_{13}})_{,12}
+ ({\tau_{31}} - {\tau_{21}})_{,23}
+ ({\tau_{12}} - {\tau_{32}})_{,13}]
\nonumber \\
& & +
{1 \over 2} 
[ ({\tau_{23}} - {\tau_{13}})_{,3}
+ ({\tau_{31}} - {\tau_{21}})_{,1}
+ ({\tau_{12}} - {\tau_{32}})_{,2}] ,
\label{eq:5}
\end{eqnarray}
and its transfer functions to instrumental noises and gravitational
waves are given in \cite{3} and \cite{5} respectively.  In
particular, $\zeta$ has a ``six-pulse response'' to gravitational
radiation, i.e.  a $\delta$-function gravitational wave signal
produces six distinct pulses in $\zeta$ \cite{5}, which are located
with relative times depending on the arrival direction of the wave and
the detector configuration.

Together with $\zeta$, three more interferometric combinations,
($\alpha, \beta, \gamma$), jointly generate the entire space of
interferometric combinations \cite{3}, \cite{5}, \cite{6}. Their
expressions in terms of the measurements $s_{ij}$, $\tau_{ij}$ are as
follows
\begin{eqnarray}
\alpha & = & s_{21} - s_{31} + s_{13,2} - s_{12,3} + s_{32,12} -
s_{23,13} + {1 \over 2} \ [({\tau_{23}} - {\tau_{13}})_{,2}
+ ({\tau_{23}} - {\tau_{13}})_{,13} + ({\tau_{31}} - {\tau_{21}})
\nonumber \\
& &
\ + \ ({\tau_{31}} - {\tau_{21}})_{,123} + ({\tau_{12}} - {\tau_{32}})_{,3}
+ ({\tau_{12}} - {\tau_{32}})_{,12}] \ ,
\label{eq:6}
\end{eqnarray}
\noindent
with $\beta$, and $\gamma$ derived by permuting the spacecraft indices
in $\alpha$. Like in the case of $\zeta$, a $\delta$-function
gravitational wave produces six pulses in $\alpha$, $\beta$, and
$\gamma$. In equations (\ref{eq:5}, \ref{eq:6}) it is important to
notice that the $\tau_{ij}$ measurements from each spacecraft always
enter into the interferometric measurements as differences taken at
the same time. This property naturally suggests a locking
configuration that makes these differences equal to zero, as we will
show in the next section.

We remind the reader that the four interferometric responses ($\alpha,
\beta, \gamma, \zeta$) satisfy the following relationship
\begin{equation}
\zeta - \zeta_{,123} =  \alpha_{,1} - \alpha_{,23} + \beta_{,2} -
\beta_{,13} + \gamma_{,3} - \gamma_{,12} \ .
\label{eq:7}
\end{equation}
Jointly they also give the expressions of the interferometric
combinations derived in \cite{3}, \cite{5}: the Unequal-arm
Michelson (${\rm X}, {\rm Y}, {\rm Z}$), the Beacon (${\rm P}, {\rm
  Q}, {\rm R}$), the Monitor (${\rm E}, {\rm F}, {\rm G}$), and the
Relay (${\rm U}, {\rm V}, {\rm W}$) responses
\begin{eqnarray}
{\rm X}_{,1} & = & \alpha_{,23} - \beta_{,2} - \gamma_{,3} + \zeta \ ,
\label{eq:8} \\
{\rm P} & = & \zeta - \alpha_{,1} \ ,
\label{eq:9} \\
{\rm E} & = & \alpha - \zeta_{,1} \ ,
\label{eq:10} \\
{\rm U} & = & \gamma_{,1} - \beta \ ,
\label{eq:11}
\end{eqnarray}
with the remaining expressions obtained from equations (\ref{eq:8},
\ref{eq:9}, \ref{eq:10}, \ref{eq:11}) by permutation of the spacecraft
indices. All these interferometric combinations have been shown to add
robustness to the mission with respect to failures of subsystems, and
potential design, implementation, or cost advantages \cite{3},
\cite{5}.
 
\subsection{Locking Conditions}

The space of all possible interferometric combinations can be
generated by properly time shifting and linearly combining the four
combinations ($\alpha, \beta, \gamma, \zeta$), as given above.
Although they have been derived by applying TDI to the twelve one-way
Doppler data, in what follows we will show that they can also be
written in terms of properly selected and time shifted two-way and
one-way Doppler measurements. These can be generated by phase locking
five of the six lasers to one of them, as it is described below.

Assume, without loss of generality, the laser on bench $1^*$ to be the
master. Although there are several other possible locking schemes, the one
chosen minimizes the number of locking conditions between the master
and any given slave. Furthermore, locking schemes relying on more than
one master could be implemented, but we will not address those in this
paper. We will also assume the spacecraft to be stationary relative to
each other. This assumption simplifies the analysis, and does not
affect the validity of the general result \cite{8}.
\begin{figure}
\centering
\includegraphics[width=5in, angle=0]{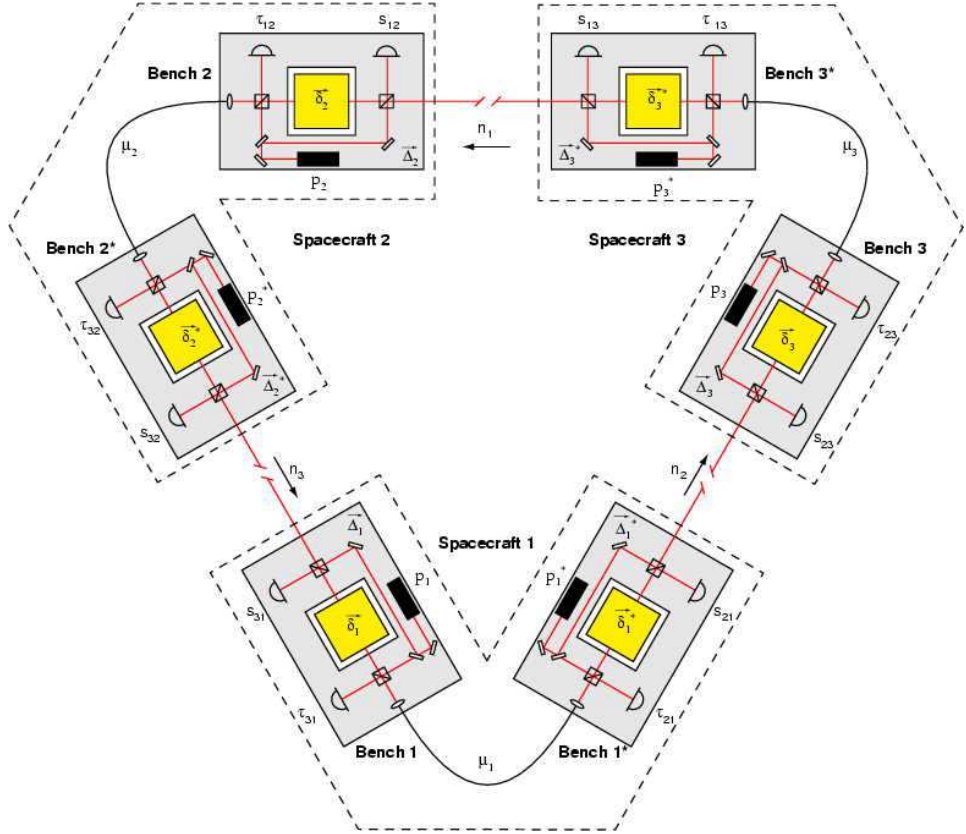}
\caption{Simplified optical layout of the LISA interferometer,
showing all the optical benches, proof masses and lasers.}
\end{figure}
Under this assumption, the frequency provided by the master laser
$1^*$ can be used as input reference for the slaved lasers. In other
words the slaves will then have the same center frequency as the master, and
their phase fluctuations will be related to the fluctuations of the
master laser as well as any other fluctuations introduced into the
received light beam prior to reception and locking.

In order to understand the topology of the beams as the various slaves
are locked to the master, let us follow the light paths from the
master laser, $1^*$, to the slaves, as shown in Figure 3.  Let us
start first with the light beam that is bounced off the back of the
proof-mass $1^*$. This beam is then directed to the other bench, $1$,
where it is used as the input frequency reference for the laser there,
and the measurement $\tau_{31}$ is made. Light is then re-transmitted
back to bench $1^*$ where the measurement $\tau_{21}$ is performed.
Since the phase of the laser $1$ is locked to that of the master, the
relative phase fluctuations $\tau_{31}$ can be adjusted as follows
\begin{equation}
\tau_{31} = \tau_{21} \ ,
\label{eq:12}
\end{equation}
where we have assumed the noise introduced by phase-locking to be
negligible (on this point see the theoretical derivations in Section
III and Appendix A). Similarly, light beams from lasers $1$ and $1^*$
are transmitted to the lasers on the benches $2^*$ and $3$
respectively, where the lasers are locked to the incoming beams.  From
benches $2^*$ and $3$ beams are transmitted to the lasers on benches
$2$ and $3^*$ respectively, where again locking is performed similarly
to what is done onboard spacecraft $1$.  Finally, along arm $1$, only
two one-way relative phase measurements can be performed as it is easy
to see.  Since for the moment we have assumed a configuration with
stationary spacecraft, the optical configuration described above can
be translated into the following {\it locking conditions} on some
relative phase fluctuation measurements
\begin{equation}
\tau_{31}  = \tau_{21} \ \ \ \ , \ \ \ \ 
\tau_{13}  =  \tau_{23} \ \ \ \ , \ \ \ \ 
\tau_{32}  =  \tau_{12} \ \ \ \ , \ \ \ \ s_{23} = s_{32} = 0 \ .
\label{eq:13b}
\end{equation}
The locking conditions define specific relationships among the phase
fluctuations from various noise sources. As an example, the conditions
\begin{equation}
\tau_{31} = \tau_{21} \ \ \ \ , \ \ \ \ s_{23} = 0 \ ,
\end{equation}
imply the following relationships among the laser phase fluctuations,
the proof-mass noises, the gravitational wave signal, and the two
bench noises onboard spacecraft $1$
\begin{eqnarray}
p^*_1 & = & p_1 + \nu_{0} \ {\hat n_3} \cdot 
({\vec \delta_1} - {\vec \Delta_1}) + \nu_{0} \ {\hat n_2} \cdot 
({\vec \delta^*_1} - {\vec \Delta^*_1}) \ , 
\label{eq:13c}
\\
0 & = & s^{\rm gw}_{23} + s^{\rm opt. \ path}_{23} + 
p^*_{1,2} - p_3 + \nu_{0} \ \left[ - \ 2 {\hat n_2} \cdot {\vec \delta_3} +
{\hat n_2} \cdot {\vec \Delta_3} + {\hat n_2} \cdot {\vec
    \Delta^*_{1,2}} \right] \ ,
\label{eq:13d}
\end{eqnarray}
with similar expressions following from the other locking conditions
given in equation (\ref{eq:13b}).

If we now substitute the locking conditions into the expressions for
($\alpha, \beta, \gamma, \zeta$), we obtain their expressions in terms
of the remaining measurements
\begin{eqnarray}
\zeta_{lo.} & = &  s_{13,3} - s_{31,1} + s_{21,1} - s_{12,2} \ ,
\label{eq:14}
\\
\alpha_{lo.} & = & s_{21} - s_{31} + s_{13,2} - s_{12,3} \ ,
\label{eq:15}
\\
\beta_{lo.} & = & - s_{12} + s_{21,3} + s_{13,23} - s_{31,12} \ ,
\label{eq:16}
\\
\gamma_{lo.} & = & s_{13} - s_{31,2} + s_{21,13} - s_{12,23} \ ,
\label{eq:17}
\end{eqnarray}
where the data ($s_{12}$, $s_{13}$) are one-way, and ($s_{21}$,
$s_{31}$) are effectively two-way Doppler measurements
due to locking.

The verification that the combinations ($\alpha_{lo.}, \beta_{lo.},
\gamma_{lo.}, \zeta_{lo.}$) exactly cancel the laser phase
fluctuations as well as the fluctuations due to the mechanical
vibrations of the optical benches, can be performed by substituting
the locked phase processes (such as \ref{eq:13c}, \ref{eq:13d}) into the
expressions for ($s_{12}, s_{13}, s_{21}, s_{31}$) (which are given by
equations (\ref{eq:1}, \ref{eq:3}) and their permutations), and by
further replacing them into equations (\ref{eq:14} - \ref{eq:17}).

The main result of implementing locking is quantitatively shown by
equations (\ref{eq:14} - \ref{eq:17}) in that the number of
measurements needed for constructing the entire space of
interferometric combinations LISA will be able to generate is smaller
by a factor of three than the number of measurements needed when only
one-way data are used.

Once ($\alpha_{lo.}, \beta_{lo.}, \gamma_{lo.}, \zeta_{lo.}$) are
constructed according to the expressions given in equations
(\ref{eq:14} - \ref{eq:17}), all the other interferometric
combinations can be derived by applying the identities given in
equations (\ref{eq:8} - \ref{eq:11}). As an example, it is
straightforward to show that equation (\ref{eq:8}) implies the
following expression for the unequal-arm Michelson combination $X_{lo.}$
\begin{equation}
X_{lo.} = [s_{21} - s_{31}] - [s_{21,33} - s_{31,22}] \ , 
\label{eq:18}
\end{equation}
which coincides with the expression for $X$, derived for the first time
in \cite{7}, in terms of the two-way Doppler measurements from the two
LISA arms.

As a final comment, we have analyzed also several locking
configurations needed when the spacecraft are moving relative to each
other. We have found that there exist techniques, when locking is
implemented, which are similar to the one analyzed in \cite{2} for
removing the noise of the onboard Ultra Stable Oscillators from the
phase measurements. The conclusions derived above for the case of
stationary spacecraft are therefore general, and an analysis
covering locking configurations with moving spacecraft
is available in \cite{8}.

\section{Phase Locking Performance}


The locking conditions given in equation (\ref{eq:13b}) reflect the
assumption that the noise due to the optical transponders is
negligible. In this section we analyze the noise added by the process
of locking the phase of the local laser to the phase of the received
light. This noise will be in addition to the optical path and USO
noises, which we consider separately.

\begin{figure}[httb]
\centering 
\includegraphics[width=7in, angle=0]{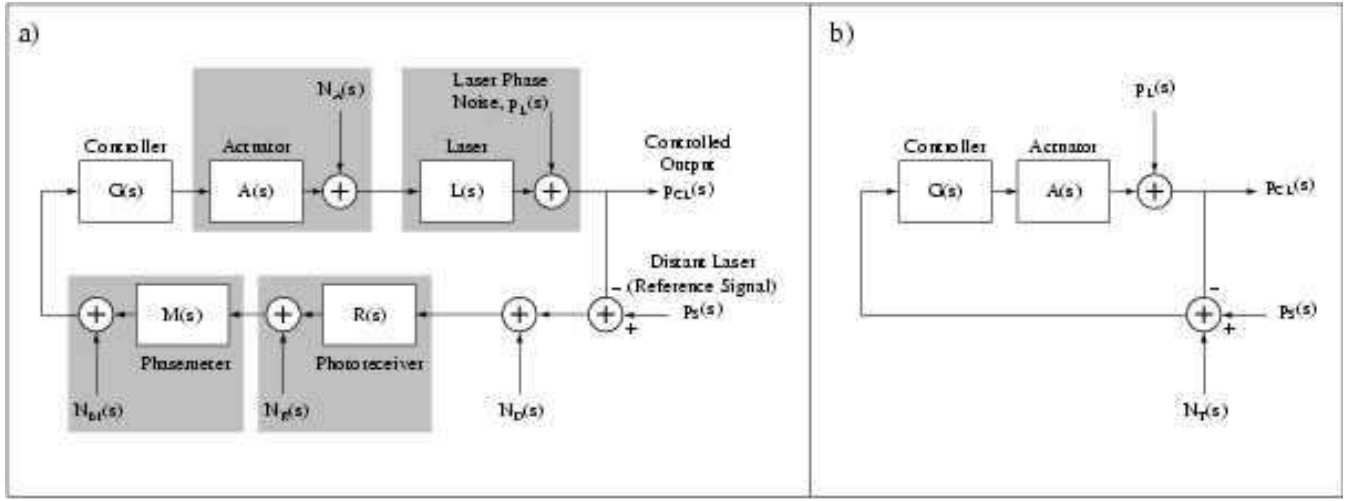}
\caption{a) Block diagram of phase locking control system showing transfer 
  functions and noise contributions of various system components.  (b)
  Simplified control system block diagram.}
\label{Servo} 
\end{figure}

A block diagram of the phase locking control system is shown in figure
\ref{Servo}(a). The system consists of a photoreceiver, phasemeter,
controller, actuator and laser (see section IV for a description of
these subsystems). Each of these subsystems can be characterized by
its transfer function and, when it applies, by a noise contribution
\cite{Abramovici}. The main inputs to the system are the phase noise
of the local laser, $p_L(s)$, and the phase fluctuations of the signal
beam from the distant spacecraft, $p_S(s)$ (with $s=\sigma + i \omega$
being the Laplace variable).  The closed loop output is the phase
noise of the retransmitted laser beam, $p_{CL}(s)$.

From the block diagram shown in figure 4a, it is easy to see that the
closed loop output phase $p_{CL}(s)$ can be written in terms of the
free-running laser phase noise, $p_L(s)$, the input signal phase
fluctuations, $p_S(s)$, and the various feedback components' transfer
functions and noises shown in Figure 4, as follows
\begin{equation}
p_{CL} = p_L+L  \left\{ N_A + A \ G \left[ N_M+M
\left(N_R+R\left\{N_D+p_S - p_{CL}\right\}\right) \right] \right\} \ .
\end{equation} 
This equation can be solved for $p_{CL}$
\begin{equation}
p_{CL} = 
{
{
p_L+L  \left\{ N_A + A \ G \left[ N_M+M
\left(N_R+R\left\{N_D+p_S\right\}\right) \right] \right\}}
\over 
{1+L \ A \ G \ M \ R}
} \ ,
\label{eq:18bis}
\end{equation}
where we have denoted with $N_R (s)$, $N_M (s)$, and $N_A (s)$ the
noises due to the photoreceiver, phasemeter, and actuator
respectively.  The detection noise, $N_D(s)$ is the error in the
measurement of the relative phase of the two beams. This noise is the
fundamental limit to the phase measurement process for the LISA
detection system (see Appendix A). Since the link from the phasemeter
to the controller will be digital, the controller noise will be due
only to the finite precision of the digital phase information. We thus
assume this noise to be negligible and do not include it in this
analysis.

Since (i) the product $M (s) \ R (s) = 1$ (the phase at the phasemeter output
ought to be equal to the phase at the input point of the
photoreceiver), (ii) the contribution of the actuator noise to the
output, $L(s) N_A(s)$, is much smaller than the noise from the
free-running laser (the free running laser noise is measured with the
actuators attached and thus intrinsically contains this noise source),
and (iii) the laser transfer function, $L(s)$, is a passive low pass
filter with a pole at several GHz (and so can be ignored in this
discussion), we conclude that the block diagram in \ref{Servo} (a) now
simplifies to that shown in figure \ref{Servo}(b), with the closed
loop output now given by
\begin{eqnarray}
p_{CL}&=&p_L+AG\left[N_T+p_S-p_{CL}\right] \ \ \ 
\longrightarrow p_{CL}=\frac{p_L+AG\left[N_T+p_S\right]}{1+AG} \\
p_{CL}-p_S&\simeq&N_T+\frac{p_L}{AG}\,\,\,\,\, (\rm{for}\,\,AG\gg1) \ .
\label{closedloop}
\end{eqnarray}
The quantity $N_T(s)$ is the total noise at the input to the
controller and is given by
\begin{eqnarray}
N_T&=&N_M+MN_R+N_D) \ .
\label{NT}
\end{eqnarray}
Equation (\ref{closedloop}) shows that phase locking will drive the
phase of the local laser, $p_{CL}(s)$, to that of the received laser,
$p_S(s)$, with an error introduced by two terms. The first term is the
total measurement noise in the phase measurement, $N_T(s)$, which is
in turn determined by the detection noise, photoreceiver noise and
phasemeter noise (see equation \ref{NT}). Recall that the detection
noise, $N_D(s)$, is the fundamental error in the measurement of the
relative phase of the two beams for the LISA detection system. A
rigorous calculation of this noise source, included in appendix
\ref{QMC}, shows that its root power spectral density is equal to
$\sim1~\mu{\rm cycle}/\sqrt{\rm Hz}$ at the output of the phase meter.
$N_R(s)$ is the electronic noise of the photoreceiver at the beat note
frequency.  Referenced to the output of the phasemeter this will be
well below the $1~\mu{\rm cycle}/\sqrt{\rm Hz}$ level. A phasemeter
noise floor of $\sim1~\mu{\rm cycle}/\sqrt{\rm Hz}$ has been set as a
requirement for the phasemeter noise. This level of performance has
already been demonstrated (\cite{8bis}, \cite{8SHA}) over the
frequency range of interest (1~mHz to 1 Hz) albeit with heterodyne
frequencies of a few kilohertz . Given these estimates of the
individual noise sources a total measurement error, $N_T (s)$ of less
than $2 \ \mu {\rm cycles}/\sqrt{\rm Hz}$ is expected.

The second source of error in Eq. (\ref{closedloop}) represents the
finite suppression of the free running laser noise.  This term is
inversely proportional to the loop gain and so can be reduced by
increasing the gain. Very high gains should be possible with LISA as
the frequencies of interest are very low \cite{Abramovici}, $\approx
10$ ~Hz and lower.  A free-running laser frequency noise of
1~MHz/$\sqrt{\rm Hz}$ at 1~mHz corresponds to a phase noise of
$10^9$~cycles/$\sqrt{\rm Hz}$.  A total loop gain of $10^{15}$ is
therefore required to suppress the contribution of laser frequency
noise down to 1~$\mu{\rm cycle}/\sqrt{\rm Hz}$. At low frequencies the
laser phase is altered by changing the temperature of the laser
crystal. This actuator has a typical (voltage to frequency) gain of
5~GHz/Volt or $5 \times10^{12}~{\rm cycles}/\sqrt{\rm Hz}$ at 1~mHz.
Thus a controller gain of 200~V/cycle is needed. This gain requirement
could be eased by ``pre-stabilizing'' the laser frequency, for example
by locking to a low finesse cavity or other frequency reference.
Initial results from bench top experiments
(\cite{McNamaraPL},\cite{Ye}) indicate that loop gains of the order of
$10^{15}$ should be achievable, and hence that pre-stabilization may
not be necessary.

The master-slave configuration phase locking requirement should be
that the noise introduced by the locking process is insignificant
compared to the $20~\mu{\rm cycle}/\sqrt{\rm Hz}$ of optical path
noise allocated in the LISA noise budget \cite{1}. As the analysis
above has shown, phase locking is limited only by the measurement
noise (assuming adequate gain). This measurement noise is common to
both the one-way and the master-slave schemes and so, given adequate
gain of the phase locking loop, there will be no difference in the
performance of the two systems.

\section{Hardware Requirements}

In this section we compare the hardware needed by the two
implementations of TDI discussed in the previous sections. The
discussion will focus on the minimum hardware requirements and will
not fully consider redundancy or fallback options.  We will consider
LISA as composed of several basic subsystems, and compare the type and
quantity of the components required by each scheme.

\subsection{Laser Frequency Stabilization System}

Both schemes rely on a sufficiently high laser frequency stability.
This is because the cancellation of laser frequency noise via TDI is
not exact due to finite accuracy of the arm length knowledge, finite
timing accuracy, imperfect clock synchronization, and sampling time
jitters. The frequency stabilization system could be composed of
either an optical cavity, a gas cell, or a combination of both. The
output is an error signal proportional to the difference between the
laser frequency and the resonance frequency of the reference. Assuming
that an optical cavity is used as the frequency reference with a
Pound-Drever-Hall locking \cite{PDH} readout, the frequency
stabilization system will consist of an electro-optic phase modulator,
an optical cavity, a photoreceiver, a double-balanced mixer, and a low
pass filter \cite{8tris}.

\subsection{Phasemeter}

A phasemeter is a device capable of measuring the phase of a
photoreceiver output relative to the local USO. One could distinguish
between two types of phasemeter: single-quadrature phasemeters,
denoted PM(S), and full-range phasemeters, denoted PM(F). An example
of a single-quadrature phasemeter is a mixer. It will have only a
limited linear range as the output is generally sinusoidal. A
single-quadrature phasemeter has several potential advantages over a
full-range phasemeter including lower noise, higher speed operation,
increased reliability and lower power consumption. However, it is
restricted in usefulness to closed loop operation only and is
therefore not suitable for use in the one-way method. A
single-quadrature phasemeter could potentially be used in the
master-slaves configuration where one laser is phase locked to another
with a fixed phase shift (see section IV. E).

A full-range phasemeter has an output that is linearly proportional to
phase over the entire range $-\pi$ to $\pi$. An example of a
full-range phasemeter is a zero-crossing time interval analyzer. As we
show in section V. E, phasemeters used in an open-loop configuration
must have a dynamic range of at least $\sim10^{10}$ at $0.1$ mHz.

\subsection{Controller}

A controller takes the signal from either a frequency stabilization
system or phasemeter, amplifies and filters it appropriately, and
feeds it back to the frequency/phase actuators of a laser. A
controller is needed to frequency lock a laser to a frequency
reference or to phase lock one laser to another. In practice the
frequency locking and phase locking controllers will differ by a pole
in the controller transfer function and by a gain factor. Depending on
the sophistication of its design, a controller could potentially be
reconfigured in-flight to perform either function.

\subsection{Photoreceivers}

Each scheme requires twelve photoreceivers to measure the interference
from the front and back of the proof masses. The photoreceiver unit
will consist of a photodiode and low-noise electronic amplifiers.
Although the photoreceivers for LISA will contain quadrant photodiodes
for alignment sensing, this is irrelevant for the following discussion
which assumes that single element photodiodes are used.

\begin{table}
 {\centering \begin{tabular}{|c|c|c|}
 \hline
Spacecraft/Bench & Master-Slave Configuration &
One-Way Method\\
 \hline
 \hline
SC $1^*$& 1 FS, 2 PR, 2 PM(F), 1 Controller&1 FS, 2 PR, 2 PM(F), 1 Controller\\
 \hline
 SC 1& 2 PR, 2 PM(F), 1 Controller&1 FS, 2 PR, 2 PM(F), 1 Controller\\
\hline
SC $2^*$&2 PR, 1 PM(F), 1 PM(S), 1 Controller&1 FS, 2 PR, 2 PM(F), 1 Controller\\
\hline
SC 2&2 PR, 2 PM(F), 1 Controller&1 FS, 2 PR, 2 PM(F), 1 Controller\\
\hline
SC $3^*$&2 PR, 2 PM(F), 1 Controller&1 FS, 2 PR, 2 PM(F), 1 Controller\\
\hline
SC 3&2 PR, 1 PM(F), 1 PM(S), 1 Controller&1 FS, 2 PR, 2 PM(F), 1 Controller\\
\hline

\end{tabular}\par}

\caption{\label{tbl:comp1} Comparison of minimum system components
required for each scheme. FS: frequency stabilization system, 
PM(S): single-quadrature phasemeter, PM(F) full-range phasemeter, 
PR: photoreceiver.}
\end{table}

\subsection{Requirements for the Master-Slave Configuration}

Table I summarizes the system components needed on each optical bench.
The quantities shown represent the minimum requirements with no
redundancy included. In the master-slave configuration scheme one
master laser is frequency stabilized to its own frequency reference.
All other lasers are phase locked in a chain to this master laser, as
described in section IIA. For this reason the minimum system
requirement is only one frequency stabilization system. However, the
capability of stabilizing other lasers to a local stabilization system
should be included for redundancy against failure of the master's
stabilizing device.  Providing each laser with a frequency
stabilization system would provide a high level of redundancy and
maintain compatibility with the one-way mode of operation.

The master-slave configuration will require at least four full-range
phasemeters for the main signal read out photoreceivers, $s_{21}$,
$s_{31}$, $s_{12}$ and $s_{13}$. Full-range phasemeters will also be
required for measuring the signals derived from the back side of the
proof masses, $\tau_{ij}$, as the lasers on adjacent benches are
locked by suppressing the difference of the phasemeter outputs,
$\tau_{ij}-\tau_{kj}=0$. If just one of these phasemeter outputs were
used for phase locking then a single-quadrature phasemeter could be
utilized. However, the noise in the optical fiber linking the benches,
$\mu_{i}$, would then be imposed on the phase of the slave laser.
Although this noise would be removed by including the second detector
in the time-delay interferometry processing, this would impose
unnecessary requirements on the stability of the fiber link to prevent
increasing the slave laser's frequency noise. Furthermore, by
suppressing the difference of these phasemeter outputs, we do not need
to record this information for processing, as only the difference of
the phasemeter outputs appears in the TDI equations (see for example,
equations \ref{eq:5} and \ref{eq:6}). Single-quadrature phasemeters
could potentially be used on the remaining two photoreceivers where
phase locking of the beams returning to spacecraft $1$ ($s_{23} =
s_{32} = 0$) is performed. This is a relatively minor simplification,
as suitable full-range phasemeters must be developed for the remaining
ten photoreceivers. If the phase locking hierarchy must be reordered,
for example due to a frequency stabilization system failure, then
full-range phasemeters will also be required at these positions.
Finally, implementing full-range phasemeters at all photoreceiver
outputs will maintain compatibility with the one-way method. For these
reasons we will drop the (F) or (S) suffix for the phasemeters and
assume that only full-range phasemeters are used.

Table II summarizes the total number of components required and the
number of data streams to be recorded by each scheme. Using the
master-slave configuration only four data streams remain to be
measured ($s_{21}$, $s_{31}$, $s_{12}$ and $s_{13}$) as all other
variables have been suppressed to effectively zero and therefore do
not need to be recorded (although they should be monitored to ensure
proper operation of the phase locking systems). One potential concern
with the master-slave configuration is in the non-local nature of the
control system, that is to say the main phase input to the control
loop comes from the light from the distant spacecraft. The amplitude
and phase of this beam could be adversely affected by many factors
such as spacecraft alignment. Although this will also affect the
quality of the one-way measurements, it could be more detrimental to
the master-slave configuration. For example, the signal intensity could
become so low to cause loss of phase lock entirely. Furthermore,
because all the slaves are linked to the master by the phase-locking
chain, if one phase-locking link is disrupted then all downstream
links may also be lost. The severity of this non-local control problem
depends on how often lock will be disrupted, and the difficulty of
lock reacquisition.

\begin{table}
 {\centering \begin{tabular}{|c|c|c|c|}
 \hline
Subsystem&Master-Slave Configuration &One-Way Method\\
 \hline
 \hline
FS&1&6\\
 \hline
 PR&12&12\\
\hline
 PM&12&12\\
\hline
Controllers&6&6\\
\hline
Observables&4&9\\
\hline
\end{tabular}\par}
\caption{\label{tbl:comp2}Summary of minimum system components
required for each scheme. FS: frequency stabilization system, 
PM: phasemeter, PR: photoreceiver, Observables: number of data 
streams required for processing.}
\end{table}

\subsection{Requirements for the One-way Method}

The one-way method employs a very symmetric configuration consisting
of three identical spacecraft each containing two identical optical
benches. The components needed on each optical bench are shown in the
right hand column of Table I. The six lasers are frequency stabilized
to their six respective frequency references. The phases of the beat
notes of each local laser with the lasers from the adjacent spacecraft
and bench are measured by full-range phasemeters to provide twelve
data streams.  The data from the phasemeters at the back of the proof
masses on adjacent benches can be combined before being recorded
without loss of generality. This reduces the total number of data
streams to be recorded and processed to nine, as shown in Table II.
However, as we will show in section V C the number of data streams to
be exchanged between spacecraft will be the same for the one-way and
master-slaves configuration.

From Tables I and II it is clear that the one-way and master-slaves
configurations are almost identical in terms of the quantity of
components required. However, there are several more subtle
differences in the hardware requirements of the two schemes. A
disadvantage of the one-way method is that the laser frequencies may
differ by as much as $\pm300$~MHz if high finesse cavities are used as
the frequency references. This maximum frequency offset is determined
by the free spectral range of the reference cavity, where a cavity
with a round trip optical path of 0.5~m has been assumed. This large
frequency offset will place greater demands on the photoreceivers'
bandwidth than the master-slave configuration where frequency offsets
can be kept to the minimum dictated by the Doppler shifts (less than
$\pm 10$~MHz). Not only does the high heterodyne frequency place strict
requirements on the photodetector bandwidth, but also on the bandwidth
stability. For example, assume that photoreceivers with a bandwidth of
$f_{bw}\approx1$~GHz are used for the main signal readouts. Although a
heterodyne frequency, $f_{h}$, of 300~MHz is within the 1~GHz
photoreceiver bandwidth there will be a 0.05~cycle phase delay at this
frequency (the phase shift is equal to -arctan$(f_{h}/f_{bw})$ radians
for a simple single-pole type frequency response). If the bandwidth of
a photodetector changes, then this phase shift will also change in a
way that is indistinguishable from the effects of a gravitational
wave. A simple calculation shows that a bandwidth change of a mere
$0.023\%$ (or 230kHz) would introduce a phase signal of
10~$\mu$cycles/$\sqrt{\rm Hz}$ for information at 300~MHz. If the
heterodyne frequency is kept to 10~MHz or less, then a photoreceiver
bandwidth change of more than $0.6\%$ (or 6~MHz) would be needed to
produce a 10~$\mu$cycles/$\sqrt{\rm Hz}$ phase shift.

In section III it was shown that a total loop gain of $10^{15}$ at 1
mHz is required to ensure that the phase locking loop performs
correctly. Using the one-way method the phase locking loops are
replaced by frequency locking to the reference cavity. The frequency
stabilization system is expected to be limited by fluctuations in
length of the cavity at a level of the order of 10 Hz/$\sqrt{\rm Hz}$
at 1 mHz.  Under this condition no advantage is gained by suppressing
the measured laser noise below this level and so a loop gain of
$\approx10^{6}$ should suffice. The one-way method therefore requires
a controller gain of only $2\times10^{-7}$~Volts/cycle at 1~mHz
compared to the 200~Volts/cycle needed for the master-slave
configuration.

\section{Accuracy and precision requirements}

The limitations on the effectiveness of the TDI technique, either when
the one-way or the master-slave configuration is implemented, come
not only from all the secondary noise sources affecting the
measurements $s_{ij} \ , \ \tau_{ij} \ , \ i,j=1, 2, 3$ (such as
proof-mass and optical path noises) but most importantly from the
finite accuracy and precision of the quantities needed to synthesize
the laser-noise free observables themselves. In order to synthesize
the four generators of the space of all interferometric combinations,
we need:

\noindent
(i) to know the distances between the three pairs of spacecraft;

\noindent
(ii) to synchronize the clocks onboard the three spacecraft, which
are used in the data acquisition and digitization process;

\noindent
(iii) to be able to apply time-delays that are not integer multiples of
the sampling time of the digitized phase measurements;

\noindent
(iv) to minimize the effects of the jitter of the
sampling times themselves; and 

\noindent
(v) to have sufficiently high dynamic range in the digitized data in
order to be able to recover the gravitational wave signal after
removing the laser noise.

In the following subsections we will assume the secondary random
processes to be due to the proof masses and the optical path noises
\cite{3}. We will estimate the minimum values of the accuracies and
precisions of the physical quantities listed above that allow the
suppression of the laser frequency fluctuations below the level
identified by the secondary noise sources. For each physical quantity
the estimate of the accuracy and/or precision needed will be performed
by assuming all the remaining errors to be equal to zero.  Our
estimates, therefore, will provide only an order of magnitude estimate
of the accuracies and/or precisions needed for successfully
implementing TDI.

\subsection{Armlength accuracy}

The TDI combinations described in the previous sections rely on the
assumption of knowing the armlengths sufficiently accurately to
suppress laser noise well below other noises. Since the three
armlengths will be known only within the accuracies $\delta L_i \ , \ 
i=1, 2, 3$ respectively, the cancellation of the laser frequency
fluctuations from the combinations ($\alpha, \beta, \gamma, \zeta$)
will no longer be exact.  In order to estimate the magnitude of the
laser fluctuations remaining in these data sets, let us define $\hat
L_i \ , \ i=1, 2, 3$ to be the estimated armlengths of LISA.  They are
related to the {\it true} armlengths $L_i \ , \ i=1, 2, 3$, and the
accuracies $\delta L_i \ , \ i=1, 2, 3$ through the following
expressions
\begin{equation}
\hat L_i = L_i + \delta L_i \ \ \ \ \ , \ \ \ \ \ i=1, 2, 3 \ .
\label{eq:19}
\end{equation}
In what follows we will treat the three armlengths $L_i \ , \ i=1, 2,
3$ as constants equal to $16.7$ light seconds. We will derive later on
the time scale during which such an assumption is valid. We will also
assume to know with infinite accuracies and precisions all the
remaining physical quantities needed to successfully synthesize the
TDI generators.

If we now substitute equation (\ref{eq:19}) into equations
(\ref{eq:5}), and expand it to first order in $\delta L_i$, it is easy
to derive the following approximate expression for ${\hat \zeta} (t)$,
which now will show a non-zero contribution from the laser noises
\begin{equation}
{\hat \zeta} (t) \simeq \zeta (t) + 
[{\dot p}_{2,13} - {\dot p}^*_{3,12}] \ \delta L_1
+ [{\dot p}_{3,12} - {\dot p}^*_{1,23}] \ \delta L_2
+ [{\dot p}_{1,23} - {\dot p}^*_{2,13}] \ \delta L_3 \ ,
\label{eq:20}
\end{equation}
where the ``${\dot \ }$'' denotes time derivative. Time-Delay Interferometry
can be considered effective if the magnitude of the remaining
fluctuations from the lasers are smaller than the fluctuations due to
the other noise sources entering in $\zeta (t)$, namely proof mass and
optical path noises. This requirement implies a limit in the
accuracies of the measured armlengths.

Let us assume the six laser phase fluctuations to be uncorrelated to
each other, their one-sided power spectral densities to be equal, the
three armlengths to differ by a few percent, and the three armlength
accuracies also to be equal. By requiring the magnitude of the
remaining laser noises to be smaller than the secondary noise sources,
it is straightforward to derive, from Eq. (\ref{eq:20}) and the
expressions for the proof mass and optical path noises entering into
$\zeta (t)$ given in \cite{3}, the following constraint on the common
armlength accuracy $| \delta L_{\zeta}|$
\begin{equation}
|\delta L_{\zeta}| \le \ {1 \over {2 \pi f}} \ 
\sqrt{ 
{4 \ \sin^2(\pi f L) \ S^{proof \ mass}_p (f) + S^{optical \
    path}_p (f)} \over {S_p (f)}
} \ .
\label{eq:21}
\end{equation}
Here $S_p$, $S^{proof \ mass}_p$, $S^{optical \ path}_p$ are the
one-sided power spectral densities of the phase fluctuations of a
stabilized laser, a single proof mass, and a single-link optical path
respectively \cite{3}. If we take them to be equal to the following
functions of the Fourier frequency $f$ \cite{2,3}
\begin{eqnarray}
S_p (f) & = & 2.3 \times 10^{-1}  \ f^{- 8/3} + 1.4 \times  10^{-9} \ f^{-27/5} \ {\rm
  cycles}^2 \ {\rm Hz}^{-1} \ ,
\label{eq:22}
\\
S^{proof \ mass}_p (f) & = & 5.8 \times 10^{-21} \ f^{-4} \ {\rm
  cycles}^2 \ {\rm Hz}^{-1} \ ,
\label{eq:23}
\\
S^{optical \ path}_p (f) & = & 4.1 \times 10^{-10} \ {\rm
  cycles}^2 \ {\rm Hz}^{-1} \ ,
\label{eq:24}
\end{eqnarray}
(where $f$ is in Hz), we find that the right-hand-side of the
inequality given by equation (\ref{eq:21}) reaches its minimum of
about $16$ meters at the Fourier frequency $f_{min} = 1.0 \ \times
10^{-4} \ {\rm Hz}$, over the assumed ($10^{-4}, 1$) Hz LISA band.
This implies that, if the armlength knowledge $| \delta L_{\zeta} |$
can be made smaller than $16$ meters, the magnitude of the residual
laser noise affecting the $\zeta$ combination will be below that
identified by the secondary noises.  This reflects the fact that the
armlength accuracy is a decreasing function of the frequency.  For
instance, at $10^{-3}$ Hz the armlength accuracy goes up by almost an
order of magnitude to about $155$ meters.

A perturbative analysis similar to the one described above can be
performed for the remaining generators ($\alpha, \beta, \gamma$). We
find that the corresponding inequality for the armlength accuracy
required for the $\alpha$ combination, $| \delta L_{\alpha} |$, is
equal to \cite{3,7}
\begin{equation}
|\delta L_{\alpha}| \le \ {1 \over {2 \pi f}} \ 
{
\sqrt{ 
{[8 \ \sin^2(3 \pi f L) + 16 \ \sin^2(\pi f L)]
\ S^{proof \ mass}_\phi (f) + 6 \ S^{optical \
    path}_\phi (f)} \over {6 \ S_p (f)}
}} \ ,
\label{eq:25}
\end{equation}
with similar inequalities also holding for $\beta$ and $\gamma$.
Equation (\ref{eq:25}) implies a minimum of the function on the
right-hand-side equal to about $31$ meters at the Fourier frequency
$f_{min} = 1.0 \ \times 10^{-4} \ {\rm Hz}$, while at $10^{-3}$ Hz
the armlength accuracy goes up to $180$ meters.

Armlength accuracies significantly smaller than the level derived
above can be achieved by implementing laser ranging measurements along
the three LISA arms \cite{1}, and we do not expect this to be a
limitation for TDI.

In relation to the accuracies derived above, it is interesting to
calculate the time scales during which the armlengths will change by
an amount equal to the accuracies themselves.  This identifies the
minimum time required before updating the armlength values
in the TDI combinations.

It has been calculated by Folkner {\it et al.} \cite{9} that the
relative longitudinal speeds between the three pairs of spacecraft,
during approximately the first year of the LISA mission, can
be written in the following approximate form
\begin{equation}
V_{i,j} (t) = V^{(0)}_{i,j} \ \sin\left({{2 \pi t} \over
T_{i,j}}\right) \qquad \qquad (i,j) = (1,2) \ ; \ (1,3) \ ; \ (2,3) \ ,
\label{eq:26}
\end{equation}
where we have denoted with $(1,2), (1,3), (2,3)$ the three possible
spacecraft pairs, $V^{(0)}_{i,j}$ is a constant velocity, and
$T_{i,j}$ is the period for the pair $(i, j)$.  In reference \cite{9}
it has also been shown that the LISA trajectory can be selected in
such a way that two of the three arms' rates of change are essentially
equal during the first year of the mission.  Following reference
\cite{9}, we will assume $V^{(0)}_{1,2} = V^{(0)}_{1,3} \ne
V^{(0)}_{2,3}$, with $V^{(0)}_{1,2} = 1 $ m/s, $V^{(0)}_{2,3} = 13 $
m/s, $T_{1,2} = T_{1,3} \approx 4 $ months, and $T_{2,3} \approx 1$
year.  From equation (\ref{eq:26}) it is easy to derive the variation
of each armlength, for example $\Delta L_3 (t)$, as a function of the
time $t$ and the time scale $\delta t$ during which it takes place
\begin{equation}
\Delta L_3 (t) = V^{(0)}_{1,2} \ \sin\left({{2 \pi t} \over
T_{1,2}}\right) \delta t \ .
\label{eq:27}
\end{equation}
Equation (\ref{eq:27}) implies that a variation in armlength $\Delta
L_3 \approx 10 \ {\rm m}$ can take place during different time scales,
depending on when during the mission this change takes place.  For
instance, if $t \ll T_{1,2}$ we find that the armlength $L_3$ changes
by more than its accuracy ($\approx 10$ meters) after a time $\delta t = 2.3
\times 10^3$ seconds.  If however $t \simeq T_{1,2}/4$, the armlength
will change by the same amount after only $ \delta t \simeq 10$ seconds
instead.

\subsection{Clock synchronization accuracy}

The effectiveness of the TDI data combinations requires the clocks
onboard the three spacecraft to be synchronized.  Since the clocks
will be synchronized with a finite accuracy, the laser noises will no
longer cancel out exactly and the fraction of the laser frequency
fluctuations that will remain into the TDI combinations will be
proportional to the magnitude of the synchronization accuracy.  In
order to identify the minimum level of off-synchronization among the
clocks that can be tolerated, we will proceed by treating one of the
three clocks (say the clock onboard spacecraft $1$) as the master
clock defining the time for LISA, and the other two to be synchronized
to it.  The relativistic (Sagnac) time-delay effect due to the fact
that the LISA trajectory is a combination of two rotations, each with
a period of one year, will have to be accounted for in the
synchronization procedure.  This is a procedure well known in the field of
time-transfer, and we refer the reader to the appropriate literature
for discussions on this point \cite{10}. Here we will disregard this
relativistic effect, and assume it can be compensated for with an
accuracy better than the actual synchronization accuracy we
derive below.

Let us denote by $ \delta t_2$, $ \delta t_3$, the time accuracies
(time-offsets) for the clocks onboard spacecraft $2$ and $3$
respectively. If $t$ is the time onboard spacecraft $1$, then what is
believed to be time $t$ onboard spacecraft $2$ and $3$ is actually
equal to the following times
\begin{eqnarray}
{\hat t_2} = t + \delta t_2 \ ,
\label{eq:28}
\\
{\hat t_3} = t + \delta t_3 \ .
\label{eq:29}
\end{eqnarray}
If we now substitute equations (\ref{eq:28}, \ref{eq:29}) into the
equation (\ref{eq:5}) for $\zeta$, for instance, and expand it to
first order in $\delta t_i \ , \ i=2, 3$, it is easy to derive the
following approximate expression for ${\hat \zeta} (t)$, which shows
the following non-zero contribution from the laser noises
\begin{equation}
{\hat \zeta} (t) \simeq  \zeta (t) + 
[{\dot p}_{1,23} - {\dot p}^*_{3,12} + {\dot p}^*_{2,13} - {\dot
  p}_{2,13}] \ \delta t_2
+ [{\dot p}_{2,13} - {\dot p}^*_{1,23} + {\dot p}^*_{3,12} - {\dot
  p}_{3,12}] \ \delta t_3 \ .
\label{eq:30}
\end{equation}
By requiring again the magnitude of the remaining fluctuations from
the lasers to be smaller than the fluctuations due to the other
(secondary) noise sources affecting $\zeta (t)$, it is possible to
derive an upper limit for the accuracies of the synchronization of the
clocks.  If we assume again the six laser phase fluctuations to be
uncorrelated to each other, their one-sided power spectral densities
to be equal, the three armlengths to differ by a few percent, and the
two time-offsets' magnitudes to be equal, by requiring the magnitude
of the remaining laser noises to be smaller than the secondary noise
sources it is easy to derive the following constraint on the time
synchronization accuracy $| \delta t_{\zeta}|$
\begin{equation}
|\delta t_{\zeta}| \le \ {1 \over {2 \pi f}} \ 
\sqrt{ 
{12 \ \sin^2(\pi f L) \ S^{proof \ mass}_p (f) + 3 \ S^{optical \
    path}_p (f)} \over {4 \ S_p (f)}
} \ ,
\label{eq:31}
\end{equation}
with $S_p$, $S^{proof \ mass}_p$, $S^{optical \ path}_p$ again as
given in equations (\ref{eq:22}-\ref{eq:24}).

We find that the right-hand-side of the inequality given by equation
(\ref{eq:31}) reaches its minimum of about $47$ nanoseconds at the
Fourier frequency $f_{min} = 1.0 \ \times 10^{-4} \ {\rm Hz}$. In other
words, clocks synchronized at a level of accuracy better than $47$
nanoseconds will imply a residual laser noise that is smaller than 
the secondary noise sources entering into the $\zeta$ combination.

An analysis similar to the one described above can be performed for the
remaining generators ($\alpha, \beta, \gamma$). For them we find that the
corresponding inequality for the accuracy in the synchronization of
the clocks is now equal to
\begin{equation}
|\delta t_{\alpha}| \le \ {1 \over {2 \pi f}} \ 
\sqrt{ 
{[4 \ \sin^2(3 \pi f L) + 8 \ \sin^2(\pi f L)]
\ S^{proof \ mass}_p (f) + 3 \ S^{optical \
    path}_p (f)} \over {4 \ S_p (f)}
} \ ,
\label{eq:32}
\end{equation}
with equal expressions holding also for $\beta$ and $\gamma$.  The
function on the right-hand-side of equation (\ref{eq:32}) has a
minimum equal to $88$ nanoseconds at the Fourier frequency $f_{min} =
1.0 \ \times 10^{-4} \ {\rm Hz}$. As for the armlength accuracies,
also the timing accuracy requirements become less stringent at higher
frequencies. At $10^{-3}$ Hz, for instance, the timing accuracy for
$\zeta$ and $\alpha, \beta, \gamma$ go up to $446$ and $500$ ns
respectively.

A $50$ ns accuracy translates into a $15$ meter armlength accuracy,
which we argued earlier to be easily achievable by the use of laser
ranging. We therefore expect the synchronization of the three
clocks to be achievable at the level derived above.

\subsection{Telemetered signals and their sampling}

To reduce the LISA-to-Earth telemetry requirements, it is expected
that the normal operational mode will not telemeter the phase time
series to the ground directly. Rather, we expect the TDI observables
to be computed at the LISA array, and then only the (relatively low
data rate) laser-noise-free combinations transmitted to Earth. Thus,
both implementations of TDI discussed in this paper (the one-way
method and the master-slave configuration) require phase measurements
data to be exchanged among the spacecraft in order to synthesize the
four generators of the space of all interferometric combinations.
Although it is clear that the master-slave configuration implies a
smaller number of measurements than that required by the one-way
method, the actual number of data that will need to be exchanged among
the spacecraft can be made to be exactly the same for both, making
their inter-spacecraft telemetry requirements identical. This can
easily be understood by rewriting the four generators ($\alpha, \beta,
\gamma, \zeta$) in the following forms
\begin{eqnarray}
\zeta & = & 
[s_{21,1}- s_{31,1}] + {1 \over 2} [
({\tau_{31}} - {\tau_{21}})_{,1}
+ ({\tau_{31}} - {\tau_{21}})_{,23}
]
\nonumber \\
& & + [s_{32,2}- s_{12,2}] + {1 \over 2} [
({\tau_{12}} - {\tau_{32}})_{,2}
+ ({\tau_{12}} - {\tau_{32}})_{,13}
]
\nonumber \\
& & + [s_{13,3}- s_{23,3}]
+ {1 \over 2} [
({\tau_{23}} - {\tau_{13}})_{,3}
+ ({\tau_{23}} - {\tau_{13}})_{,12}
] \ ,
\label{eq:33}
\end{eqnarray}
\begin{eqnarray}
\alpha & = & [s_{21} - s_{31}] + {1 \over 2} \ [({\tau_{31}} - {\tau_{21}})
+ ({\tau_{31}} - {\tau_{21}})_{,123}]
\nonumber \\
& & + [s_{32,12} - s_{12,3}] + {1 \over 2} \ [({\tau_{12}} - {\tau_{32}})_{,3}
+ ({\tau_{12}} - {\tau_{32}})_{,12}]
\nonumber \\
& & + [s_{13,2} - s_{23,13}] + {1 \over 2} \ [({\tau_{23}} - {\tau_{13}})_{,2}
+ ({\tau_{23}} - {\tau_{13}})_{,13}] \ .
\label{eq:34}
\end{eqnarray}
Equations (\ref{eq:33}, \ref{eq:34}) show that each generator can be
formed by summing three different linear combinations of the data,
each involving phase measurements performed onboard only a specific
spacecraft. As an example, let us assume without loss of generality
that $\zeta$ will be synthesized onboard spacecraft $1$. This means
that spacecraft $2$ and $3$ will simply need to telemeter to spacecraft
$1$ the particular combinations of the measurements they have made,
which enter into the $\zeta$ combination. Since the space of all the
interferometric combinations can be constructed by using four
generators ($\alpha, \beta, \gamma, \zeta$) we conclude that
spacecraft $2$ and $3$ will each have to telemeter to spacecraft $1$
four uniquely defined combinations of the measurements they have
performed.

The time-delay interferometric combinations require use of phase
measurements that are time-shifted with enough accuracy to bring the
laser phase noise below the secondary noise sources.  The required
time resolution in the time-shifts should be equal to about $\sim 50$
ns for shifts tens of seconds in size.  This is because the correct
sample of the shifted data should be as accurate as the armlength
accuracy itself.  It can be shown that performing the time-shifting,
on data sampled at $\sim 10$ Hz, by using digital interpolation
filters, does not provide the required accuracy to effectively cancel
the laser phase noises (see Appendix B for a detailed calculation).

An alternative approach \cite{11} for achieving a timing accuracy of
at least $50$ ns would be to sample each measurement at $\sim 20$ MHz
or higher and store $\sim 2.0 \times 10^9$ samples in a ring buffer
for obtaining the data points of this measurement at the needed times.
The phasemeter would then average these measurements over a fixed time
period (perhaps a tenth of a second) centered around the sampled times
at which the phase measurements are needed.  The data is then
exchanged among the spacecraft and the TDI combinations are formed.

This method can however be further refined by actually sampling every
phase difference a few times, each time at $\sim 10$ Hz, but with a
delay between the start time of every sampled version of the same
phase difference. That is, we envision triggering the phasemeter such
that the time series are sampled at the times required to form the TDI
combinations. In this case the limitation of the finite sampling
time in the determination of the delayed phase measurement is replaced
by the timing precision of the phase measurements, which can be many
orders of magnitude smaller than the smallest sampling time of the
phasemeter, as we will show below. 

As a concrete example, the ($\alpha_{lo.}, \beta_{lo.}, \gamma_{lo.},
\zeta_{lo.}$) basis in the master-slave configuration requires
measurements $\{s_{21}, s_{21,1}, s_{21,3}, s_{21,13}, s_{31},
s_{31,1}, s_{31,2}, s_{31,12}\}$ from spacecraft 1, measurements $\{
s_{12}, s_{12,2}, s_{12,3}, s_{12,23} \}$ from spacecraft 2, and
measurements $\{s_{13}, s_{13,2}, s_{13,3}, s_{13,23}\}$ from
spacecraft 3. By sampling the data $s_{21}$ at the times $n / f_s, n /
f_s - L_1, n / f_s - L_3, n / f_s - L_1 - L_3$, where $f_s \sim 10$ Hz
is the sampling frequency, and $n = 0, 1, 2, ...$ (and similarly for
$s_{12}, s_{13}$ and $s_{31}$) we can obtain the entire data at the
required times, these being limited only by the timing precision of
the phasemeters, the time synchronization accuracies of the clocks,
and the armlength accuracies. This scheme requires sampling each
signal four times at a sampling frequency ($10$ Hz) much smaller than
what would be needed if sampling the data to the granularity required
by the TDI combinations. Of course to correctly sample at $10$ Hz we
must first ensure that the signal frequency bandwidth is less than $5$
Hz to avoid aliasing problems.

In practice, the times at which every sample from a given signal are
taken could be adjusted every $1/f_s$, in order to protect the quality
of the laser noise cancellation against drifting armlengths.  This
requires an adequate model of the spacecraft orbits, which could be
updated as needed from spacecraft ranging data.

\subsection{Sampling time jitter}

The sampling times of all the measurements needed for synthesizing the
TDI combinations will not be constant, due to the intrinsic timing
jitters of the digitizing systems (USOs and phasemeters). Within the
digitizing system, the USO is expected to be the dominant source of
time jittering in the sampled data. Presently existing, space
qualified, USO can achieve an Allan standard deviation of about
$10^{-13}$ for integration times from $1$ to $10000$ seconds.  This
timing stability translates into a time jitter of about $10^{-13}$
seconds over a period of $1$ second. A perturbative analysis including
the three sampling time jitters due to the three clocks shows that any
laser phase fluctuations remaining in the four TDI generators will
also be proportional to the sampling time jitters. Since the latter
are approximately four orders of magnitude smaller than the armlength
and clocks synchronization accuracies derived earlier, we conclude
that the magnitude of laser noise residual into the TDI combinations
due to the sampling time jitters can be made negligible.

\subsection{Data digitization and bit-accuracy requirement}

As shown in figure 5, the maximum of the ratio of the laser noise and
of the secondary noises phase fluctuation amplitudes occurs at the
lower end of the LISA bandwidth, and is $\sim 10^{10}$ at 0.1 mHz.
This corresponds to the minimum dynamic range for the phasemeters to
correctly measure the laser fluctuations and the weaker signals
simultaneously. An additional safety factor of $\sim 10$ should be
sufficient to avoid saturation if the noises are well described by
Gaussian statistics.

\begin{figure}[httb]
\centering 
\includegraphics[width=6in, angle=0]{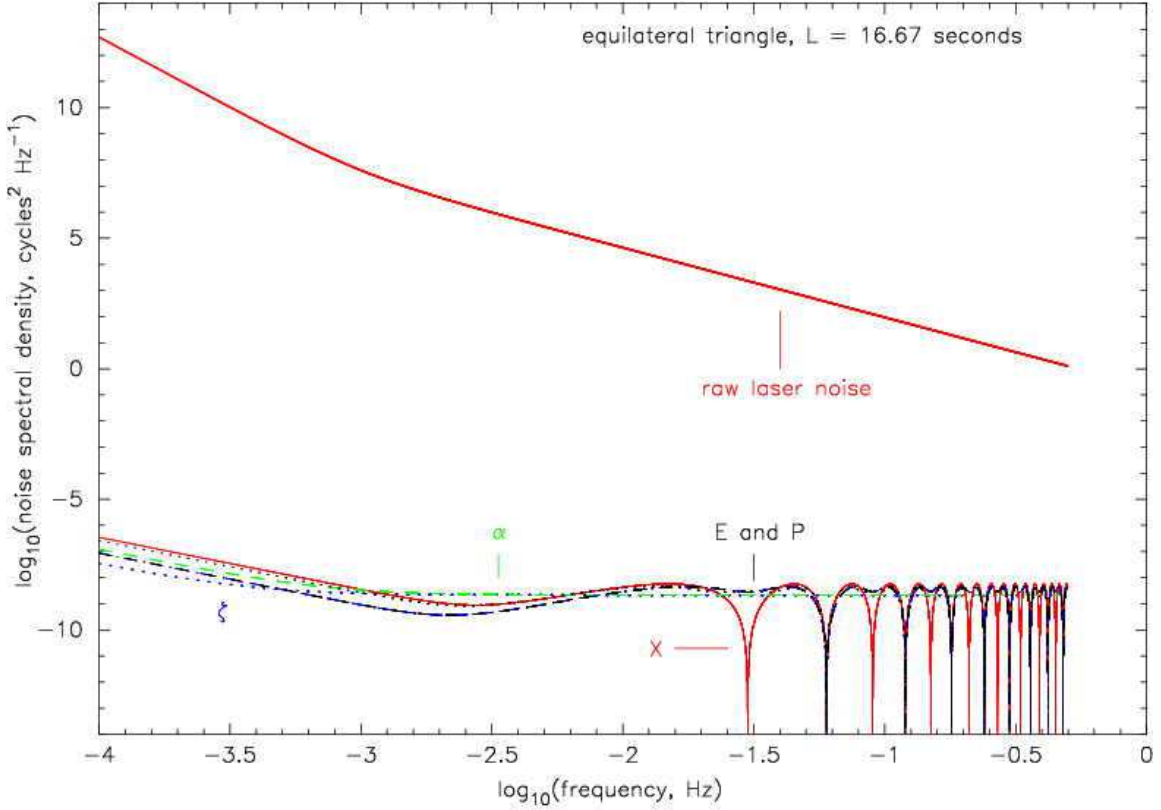}
\caption{Phase fluctuations spectra are plotted versus Fourier
  frequency for: (upper curve) raw laser noise having spectral density
  $2.3 \times 10^{-1} \ f^{- 8/3} + 1.4 \times 10^{-9} \ f^{-27/5} \ 
  {\rm cycles}^2 \ {\rm Hz}^{-1}$, and (lower curves) residual noises
  entering into the various TDI combinations.  The armlength has been
  assumed to be equal to $ L = 16.67 \ {\rm sec}$.}
\label{DR} 
\end{figure}

In terms of requirements on the digital signal processing subsystem,
this dynamic range implies that approximately $36$ bits are needed
when combining the signals in TDI, only to bridge the gap between
laser frequency noise and the other noises and gravitational wave
signals. More bits might be necessary to provide enough information to
efficiently filter the data when extracting weak gravitational wave
signals embedded into noise.

\section{Summary and Conclusions}

A comparative analysis of different schemes for implementing
Time-Delay Interferometry with LISA has been presented. In particular,
we have shown that the master-slave configuration is capable of
generating the entire space of interferometric combinations identical
to that derived by using the one-way scheme. This was done under the
assumption that the noise from the optical transponders was
negligible. Our analysis of the phase-locking control systems forming
the optical transponders shows that this is a valid assumption,
indicating that the noise introduced can be expected to be $\approx 1
\ \mu {\rm cycle}/\sqrt{\rm Hz}$.

A comparison of the hardware required for each scheme shows that the
subsystems needed are almost identical, with the only difference being
the number of frequency stabilization systems. This difference is
perhaps not significant when redundancy options are considered.  The
main disadvantage of the one-way method is that the laser frequencies
might be offset by several hundred megahertz, given the currently
envisioned optical-cavity-based frequency stabilization systems. This
places challenging constraints on the photoreceiver bandwidth and
bandwidth stability. On the other hand, the master-slave configuration
has no such problem, allowing the beat-note on the photoreceiver to be
the minimum determined by the Doppler shift.  However, there may be
concerns with the non-local nature of the phase locking system, since
its performance could be influenced, for example, by pointing
stability (which also has implications on lock acquisition). Further
studies on these issues should be performed.

Given the similarities between the two schemes, in principle either
operational mode could be implemented without major implications on
the hardware configuration. Ultimately detailed engineering studies
will identify the preferred approach.

A derivation of the armlengths and clocks synchronization accuracies,
as well as a determination of the precision requirement on the
sampling time jitter, have also been derived. We found that an
armlength accuracy of about 16 meters, a synchronization accuracy of
about $50$ ns, and the time jitter due to a presently existing Ultra
Stable Oscillator will allow the suppression of the frequency
fluctuations of the lasers below the level identified by the secondary
noise sources. A new procedure for sampling the data in such a way to
avoid the problem of having time shifts that are not integer multiples
of the sampling time was also presented, addressing one of the concerns
about the implementation of Time-Delay Interferometry.

\section*{Acknowledgments}

We would like to thank Dr. Alex Abramovici for useful discussions on
phase locking control systems, and Dr. Frank B. Estabrook for several
stimulating conversations. This research was performed at the Jet
Propulsion Laboratory, California Institute of Technology, under
contract with the National Aeronautics and Space Administration.

\appendix
\section{Fundamental limit of the LISA phase transponder}
\label{QMC}

The following is a derivation of the noise added by the optical phase
measurement. This calculation only considers noise added by the
detection system, it does not include the 20~$\mu$cycles$/\sqrt{\rm
  Hz}$ due to the optical-path noise, and it is purely quantum
mechanical \cite{Bachor}. The reason for performing the calculation in
quantum mechanical terms is to highlight that the measurement process
itself does not add shot noise and that, in principle, a perfect phase
measurement of the field could be made.

\begin{figure}[htb]
\vspace{.2in}
\centerline {
\includegraphics[width=4in]{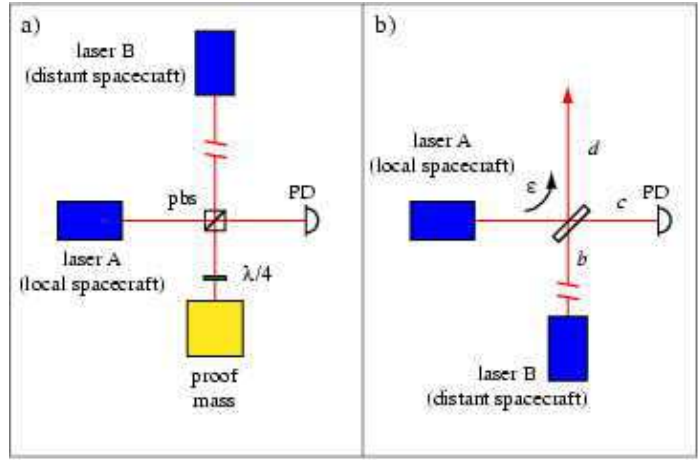}
}
\caption{(a) Simplified optical arrangement for phase locking in
  LISA. (b) Equivalent optical layout for modeling purposes.
  $\epsilon$ is the intensity reflectivity of the beam splitter.}
\vspace{.2in}
\label{diag1}
\end{figure}

Let us consider the optical configuration shown in figure
\ref{diag1}(b), which is equivalent to the optical arrangement for the
LISA interferometer (figure \ref{diag1}(a)). Let the annihilation
operators for the local and distant lasers be $\hat{a}$ and $\hat{b}$
respectively. We can represent these operators as the sum of an
average (complex number) component and an operator component
representing the field fluctuations of zero mean values:
\begin{eqnarray}
\hat{a}&=&\alpha+\delta\hat{a} \ ,
\label{defa1}
\\
\hat{b}&=&\beta+\delta\hat{b} \ .
\label{defb1}
\end{eqnarray}
In Eqs. (\ref{defa1}, \ref{defb1}) we define $\alpha$ and $\beta$ to
be real numbers. We will assume throughout these calculations that
$\delta\hat{a},\delta\hat{b}\ll\alpha,\beta$ and so terms that are of
second order in these quantities will be ignored. This approximation
holds even for the low intensities found in the LISA interferometer.

In an offset phase locking system the field $\hat{b}$ is offset from
field $\hat{a}$ by a radial frequency $\omega_{b}$. In this case
equations \ref{defa1} and \ref{defb1} become
\begin{eqnarray}
\hat{a}&=&\alpha+\delta\hat{a} \label{defa2}\\
\hat{b}&=&\beta{\rm e}^{i\omega_{b} t}+\delta\hat{b}{\rm
  e}^{i\omega_{b} t} \label{defb2} \ .
\end{eqnarray}
The annihilation operator for the field at the photoreceiver,
$\hat{c}$, will contain some fraction of $\hat{a}$ and $\hat{b}$,
\begin{eqnarray}
\hat{c}&=&\sqrt{1-\epsilon}\,\hat{a}+\sqrt{\epsilon}\,\hat{b} \ .
\end{eqnarray}

The photoreceiver measures a quantity proportional to the photon
number of this field, $n_c=\hat{c}^\dagger\hat{c}$, 
\begin{eqnarray}
\hat{c}^\dagger\hat{c}
&=&(1-\epsilon) \left[\alpha^2+\alpha(\delta\hat{a}+\delta\hat{a}^\dagger)\right] 
+\epsilon\left[\beta^2+\beta(\delta\hat{b}+\delta\hat{b}^\dagger)\right]   \\
&&
+\sqrt{\epsilon}\sqrt{1-\epsilon}\left[\alpha\beta({\rm e}^{i\omega_{b} t}+{\rm e}^{-i\omega_{b} t}) 
+\alpha(\delta\hat{b}{\rm e}^{i\omega_{b} t}+\delta\hat{b}^\dagger{\rm e}^{-i\omega_{b} t})
+\beta(\delta\hat{a}{\rm e}^{-i\omega_{b} t}+\delta\hat{a}^\dagger{\rm e}^{i\omega_{b} t})\right] \nonumber\\
&=&
(1-\epsilon) \left[\alpha^2+\alpha(\delta\hat{a}+\delta\hat{a}^\dagger)\right] 
+\epsilon\left[\beta^2+\beta(\delta\hat{b}+\delta\hat{b}^\dagger)\right] 
\nonumber  \\
&&
+2\sqrt{\epsilon}\sqrt{1-\epsilon}\Big[ (\alpha\beta +
\alpha(\delta\hat{b}+\delta\hat{b}^\dagger)+\beta(\delta\hat{a}+\delta\hat{a}^\dagger))\cos(\omega_{b} t) \nonumber \\
&&+
(\alpha(i\delta\hat{b}-i\delta\hat{b}^\dagger)-\beta(i\delta\hat{a}-i\delta\hat{a}^\dagger))\sin(\omega_{b}
t)\Big] \ . 
\label{complex1}
\end{eqnarray}
To simplify the notation, we define the quadrature operators which
represent the fluctuations in the amplitude ($\delta\hat{X}^{+}$) and
phase ($\delta\hat{X}^{-}$) quadratures of the operators $\hat{a}$ and
$\hat{b}$,
\begin{eqnarray}
\delta\hat{X}^{+}_{a}&=&(\delta\hat{a}+\delta\hat{a}^\dagger) \ , \\
\delta\hat{X}^{+}_{b}&=&(\delta\hat{b}+\delta\hat{b}^\dagger) \ ,  \\
\delta\hat{X}^{-}_{a}&=&i(\delta\hat{a}-\delta\hat{a}^\dagger) \ ,  \\
\delta\hat{X}^{-}_{b}&=&i(\delta\hat{b}-\delta\hat{b}^\dagger) \ .
\end{eqnarray}
Each of these quantities is an observable of unit variance for a
coherent state (idealized laser). Substituting these expressions into
equation \ref{complex1} we obtain,
\begin{eqnarray}
\hat{c}^\dagger\hat{c}&=&
(1-\epsilon)\left[\alpha^2+\alpha\delta\hat{X}^{+}_{a}\right]+\epsilon\left[\beta^2+\beta\delta\hat{X}^{+}_{b}\right] 
\nonumber \\
&&+ 2\sqrt{\epsilon}\sqrt{1-\epsilon}\left[(\alpha\beta 
+\alpha\delta{X}^{+}_{b}+\beta\delta{X}^{+}_{a})\cos(\omega_{b}
t)+(\alpha\delta{X}^{-}_{b}-\beta\delta{X}^{-}_{a})\sin(\omega_{b}
t)\right] \ . 
\end{eqnarray}
This equation contains three terms which can now be identified. The
first two terms arise from the intensities of fields $\hat{a}$ and
$\hat{b}$ respectively. These terms are non-interferometric in nature
and contain the intensity fluctuations of the individual input beams
scaled by the efficiency of coupling to the photoreceiver (beam
splitter ratio). The third term represents the interference between
the two fields and provides a beat note at frequency $\omega_{b}$, the
difference frequency of the two fields. This beat note itself has two
parts, an intensity noise part oscillating as $\cos(\omega_b t)$, and
a phase difference part oscillating as $\sin(\omega_b t)$.

The phase difference can be obtained, for example, by using a mixer to
demodulate the beat note down to zero. Mathematically, this is a
multiplication by $\sin(\omega_{b} t)$. Terms with a cosine multiplier
will only exhibit higher harmonics whereas terms with a sine
multiplier will mixed down to base band frequencies.

After low-pass filtering the mixer output, we obtain the following
expression for the error signal
\begin{eqnarray}
V_{e}& \equiv & \frac{(1-\epsilon)}{2}   \alpha\delta\hat{X}^{+}_{a,\omega_{b}} 
+ \frac{\epsilon}{2} \beta\delta\hat{X}^{+}_{b,\omega_{b}}
+
\sqrt{\epsilon}\sqrt{1-\epsilon}(\alpha\delta{X}^{-}_{b}-\beta\delta{X}^{-}_{a})
\ ,
\end{eqnarray}
where we have ignored the noise of the oscillator used in the mixing
process. The quantities $\hat{X}^{+}_{a,\omega_{b}}$ and
$\hat{X}^{+}_{b,\omega_{b}}$ are the amplitude quadrature operators
evaluated at the offset frequency. The factor of 1/2 in the
coefficients of these terms arises as we have only taken the sine
component of the intensity noise at the heterodyne frequency. For
heterodyne frequencies of 10~MHz and greater the intensity noise is
shot noise limited thus
$\langle(\hat{X}^{+}_{a,\omega_{b}})^2\rangle\approx
\langle(\hat{X}^{+}_{b,\omega_{b}})^2\rangle\approx 1$. The important
point is that the error signal is proportional to the intensity noise
of each beam and the relative phase noise of the two lasers.

Assuming perfect phase locking, the error signal $V_e$ is driven to
zero by actuating on the phase of $\hat{a}$. Setting $V_e=0$ we find
that the controller will attempt to force the phase quadrature
fluctuations to be,
\begin{eqnarray}
\delta{X}^{-}_{a}&=&\frac{\alpha}{\beta}\delta{X}^{-}_{b}+\frac{\sqrt{1-\epsilon}  
\alpha\delta\hat{X}^{+}_{a,\omega_{b}}}{2 \sqrt{\epsilon}\beta}  
+ \frac{\sqrt{\epsilon}\delta\hat{X}^{+}_{b,\omega_{b}}}{2
  \sqrt{1-\epsilon}} \label{perfectPL} \ .
\end{eqnarray}

Ultimately, what is of interest is the difference between the phases
of the fields $\hat{d}$ and $\hat{a}$. The annihilation operator for
the outgoing beam, $\hat{d}$, is also made up of a linear combination
of $\hat{a}$ and $\hat{b}$
\begin{eqnarray}
\hat{d}&=&\sqrt{\epsilon}\,\hat{a}-\sqrt{1-\epsilon}\,\hat{b} \ ,
\end{eqnarray}
which implies the following phase quadrature fluctuations of $\hat{d}$, $\delta
\hat{X}^-_d$
\begin{eqnarray}
\delta X^-_d&=&\sqrt{\epsilon}\delta \hat{X}^-_a
-\sqrt{1-\epsilon}\delta \hat{X}^-_b \ .
\end{eqnarray}
Under the operational configuration of perfect phase locking, we can
substitute into the equation above the expression for $\delta{X}^{-}_{a}$ given in equation
\ref{perfectPL}
\begin{eqnarray}
\delta
X^-_d&=&(\sqrt{\epsilon}\frac{\alpha}{\beta}-\sqrt{1-\epsilon})\delta\hat{X}^{-}_{b}+\frac{\sqrt{1-\epsilon}
  \alpha\delta\hat{X}^{+}_{a,\omega_{b}}}{2 \beta}
+\frac{\epsilon\delta\hat{X}^{+}_{b,\omega_{b}}}{2\sqrt{1-\epsilon}} \ .
\label{complex}
\end{eqnarray}
Since the power of the laser A (proportional to $\alpha^2$) will be a
factor of $\approx10^{8}$ larger than the power of laser B (
proportional to $\beta^2$), and also that $\langle(\delta
\hat{X}^+_b)^2\rangle\sim 1$, i.e.  the signal laser intensity is
approximately shot noise limited at the heterodyne frequency, we find
that Eq. (\ref{complex}) can be approximated as follows
\begin{eqnarray}
\delta X^-_d&\simeq&\frac{\alpha}{\beta}\left[\sqrt{\epsilon}\delta
  \hat{X}^-_b+\frac{\sqrt{1-\epsilon}\delta
    \hat{X}^+_{a,\omega_{b}}}{2}\right] \ .
\end{eqnarray}

The quantity of interest is the phase fluctuations in radians. To
compare the phase fluctuations between the two fields we need to
normalize the phase quadrature operator by the square root of the
average photon number. For $\hat{d}$, this means dividing by
$\sqrt{\epsilon}\alpha$
\begin{eqnarray}
\delta\phi_b&=&\frac{\delta\hat{X}^-_b}{\beta} \ , \\
 \delta\phi_d&=&\frac{\delta\hat{X}^-_b}{\beta}
 +\frac{\sqrt{1-\epsilon}}{2\sqrt{\epsilon}}\frac{\delta
 \hat{X}^+_{a,\omega_{b}}}{\beta} \ .
\end{eqnarray}
Thus the phase of the incoming beam will differ from the phase of the
outgoing beam by the following amount
\begin{eqnarray}
 \delta\phi_d-\delta\phi_b&=&\frac{\sqrt{1-\epsilon}}
{2\sqrt{\epsilon}}\frac{\delta \hat{X}^+_{a,\omega_{b}}}
{\beta} \ ,
\end{eqnarray}
while the root-mean-squared value of the phase error can be written
as
\begin{eqnarray}
\sigma_\phi=\sqrt{\frac{1-\epsilon}{4\epsilon}\frac{\langle|\delta\hat{X}^+_{a,\omega_b}|^2\rangle}{\bar{n}_b}}
\ .
\end{eqnarray}
Here $\bar{n}_b = \beta^2$ is the average number of photons in the
weak signal laser, and $\langle|\delta\hat{X}^+_{a,\omega_b}|^2
\rangle$ is the variance of the local oscillator intensity
fluctuations relative to the variance of quantum noise. Thus the error
depends on two parameters, the beam splitter ratio and the intensity
noise of the local oscillator. Assuming the local laser intensity is
shot noise limited at the modulation frequency and a beam splitter
ratio of approximately 100:1 ($\epsilon=0.99$) gives a phase error
with a standard deviation, $\sigma_\phi$, of approximately $5\%$ of
the shot noise limit or less than 1~$\mu$cycle$/\sqrt{\rm Hz}$ error.
This error is therefore much less than the 20~$\mu$cycles$/\sqrt{\rm
  Hz}$ optical path noise.

If necessary, this source of error could be removed by altering the
detection system to add a second detector allowing subtraction and
consequent cancellation of the intensity noise of the local oscillator
laser. Cancellation factors of 100 are readily achievable, albeit with
a slight increase in system complexity, effectively removing this
error contribution entirely.

\section{Interpolation error for generating shifted data points}

Let $g(t)$ be the true signal.  It is sampled at intervals $\Delta t =
1/f_s$, where $f_s$ is the sampling frequency, to produce the discrete
data $g[n] = g(t + n / f_s)$, $n=...,-1,0,1,...$.  Consider the
problem of estimating $g(t)$ at some time which falls in between two
sampling times, i.e. at time $t$ where $0 < |t - n_0/f_s| < \Delta t$,
for $n_0$ the value of $t/\Delta t$ rounded to the nearest integer.
It is well known that for an infinitely long dataset, this estimation
can be done without error using the Shannon formula \cite{12},
assuming that the signal has zero power above the Nyquist frequency
($f_s/2$).  The error in the estimation with a finite digital filter
can be approximated using a truncated version of the Shannon formula.
This digital filter might not be the best one for all signals, but it
should be sufficiently close to the optimal filter.  The estimated
function is given by
\begin{equation}
g_N(t) = \sum_{n=-N}^N g[n + n_0] \;{\rm sinc}(f_s t - n - n_0) \ ,
\end{equation}
for a digital filter of length $2N+1$, and where $\;{\rm
  sinc}(x)=\sin(\pi x)/\pi x$. The estimation error is $e_N(t) = g(t)
- g_N(t)$.

Changing the sampling frequency will not improve the function
estimator for a fixed digital filter size: a lower sampling frequency
would be insufficient to represent the high frequency signal
components, and a higher sampling frequency would reduce the size of
the interval over which the function is sampled to build the
estimator.

Assuming $g(t)$ to be a wide sense stationary stochastic process
\cite{13} with autocorrelation function $R_g(|t_1-t_2|) = R_g(t_1,t_2)
= E[g(t_1) g(t_2)]$ ($E[]$ denotes the expectation value), it follows
that the autocorrelation function of the estimation error is equal to
\begin{eqnarray}
R_{e_N}(t_1,t_2) = R_g(t_1 - t_2) + \nonumber \\
\sum_{m,n=-N}^N R_g\left(\frac{n+n_1-m-n_2}{f_s}\right)\;{\rm
  sinc}(f_s t_1 - n - n_1)
\;{\rm sinc}(f_s t_2 - m - n_2) \nonumber \\
- \sum_{n=-N}^N R_g(t_1 - n/f_s - n_2/f_s)\;{\rm sinc}(f_s t_2 - n - n_2) \nonumber \\
- \sum_{n=-N}^N R_g(t_2 - n/f_s - n_2/f_s)\;{\rm sinc}(f_s t_1 - n - n_1),
\end{eqnarray}
where $n_1$ ($n_2$) is the value of $t_1/\Delta t$ ($t_2/\Delta t$)
rounded to the nearest integer.  This equation shows that $e_N(t)$ is
not wide sense stationary; however, it is wide sense cyclo-stationary,
since $R_{e_N}(t_1,t_2) = R_{e_N}(t_1 + m \Delta t, t_2 + m \Delta t)$
for every integer $m$.  This is just a consequence of the error
varying quasi-periodically with the interpolation time; it is zero
when $t$ matches a sample time, maximum at the middle between two
sample times, etc.

A fair estimate of the estimation error magnitude can be obtained by
considering the stochastic process $\bar{e}_N(t) = e_N(t + \theta)$,
where $\theta$ is a random variable uniformly distributed in $[0,
\Delta t]$.  $\bar{e}_N(t)$ is wide sense stationary, and its
autocorrelation function is \cite{13}
\begin{equation}
R_{\bar{e}_N}(\tau) = f_s \int_0^{\Delta t} R_{e_N}(t+\tau,t) dt \ .
\end{equation}
The Fourier transform of $R_{\bar{e}_N}(\tau)$ gives an estimate of
the spectrum of the noise induced by the digital filters interpolation
errors.  In particular, $R_{\bar{e}_N}(0)$ is the broadband standard
deviation of the noise.

Taking $g(t)$ to be a laser phase noise with a power spectral density
that scales like $1/f^2$, and restricting attention to the frequency
range 0.1 mHz $< f <$ 1 Hz, one can calculate numerically that
$R_{\bar{e}_N}(0) / R_g(0) = 9\times 10^{-7}$ for $N=10$. Therefore, a
filter with $N=10$ is good enough only to produce a broadband error on
the shifted time series that is $\sim 3$ orders of magnitude smaller
than the laser phase noise amplitude. Changing $N$ in the numerical
integrations shows that $R_{\bar{e}_N}(0)$ scales roughly like $1/N$.
This implies that it would be impossible to use digital filters on a
slowly sampled time series to achieve the levels of noise cancellation
required by the TDI combinations.

\end{document}